\documentclass[10pt,journal,twocolumn]{IEEEtran}

\usepackage{jabbrv}

\usepackage{cite}

\ifCLASSINFOpdf
  \usepackage[pdftex]{graphicx}
  \graphicspath{{../pdf/}{../jpeg/}}
  \DeclareGraphicsExtensions{.pdf,.jpeg,.png}
\else
\fi

\usepackage{amsmath,amsfonts,amssymb,bbm}

\usepackage{array}

\ifCLASSOPTIONcompsoc
\usepackage[caption=false,font=normalsize,labelfont=sf,textfont=sf]{subfig}
\else
\usepackage[caption=false,font=footnotesize]{subfig}
\fi

\usepackage{dblfloatfix}

\usepackage{url}

\usepackage{xspace}

\usepackage{xcolor}

\usepackage{enumitem}

\usepackage{hyperref}       
\usepackage[all]{hypcap}

\usepackage{textcomp}
\renewcommand{\textrightarrow}{$\rightarrow$}

\DeclareMathOperator{\exo}{\mathrm{e}}

\newcommand{\re}{\mathrm{re}}
\newcommand{\im}{\mathrm{im}}
\newcommand{\E}{\mathbb{E}}
\newcommand{\neff}{N_{\mathrm{eff}}}
\newcommand{\cov}[1]{\mathbb{C}\textsc{ov}\left\{ #1 \right\}}
\newcommand{\R}{\mathbb{R}}
\newcommand{\Na}{\mathbbm{N}}

\newcommand{\mT}{\mathcal{T}}

\newcommand{\bfm}{{\mathbf m}}

\newcommand{\Z}{\mathbb{Z}}

\newcommand{\Om}{\Omega}

\newcommand{\bfx}{{\mathbf{x}}}

\newcommand{\bfA}{{\mathbf{A}}}

\newcommand{\bfC}{{\mathbf{C}}}

\newcommand{\bfX}{{\mathbf{X}}}
\newcommand{\hC}{\hat{C}}
\newcommand{\hr}{\hat{r}}
\newcommand{\hrho}{\hat{\rho}}

\newcommand{\tij}{T_{i \to j}}
\newcommand{\tijz}{{\mathcal T}_{i \to j}}
\newcommand{\tjiz}{{\mathcal T}_{j \to i}}

\newcommand{\htij}{\hat{T}_{i \to j}}
\newcommand{\htji}{\hat{T}_{j \to i}}

\newcommand{\htto}{\hat{T}_{2 \to 1}}

\newcommand{\Oo}{\mathcal{O}}

\newcommand{\ben}{\begin{enumerate}}
\newcommand{\een}{\end{enumerate}}
\newcommand{\beq}{\begin{equation}}
\newcommand{\eeq}{\end{equation}}
\newcommand{\noi}{\noindent}

\newcommand{\delt} {{\delta t}}

\newcommand{\di}{\mathrm{d}}
\newcommand{\var}[1]{\,\mathbb{V}\textrm{ar}[{#1}]}

\newcommand{\wm}{\omega}

\newtheorem{defi}{Definition}
\newtheorem{theorem}{{\bf Theorem}}[section]
\newtheorem{lemma}{{\bf Lemma}}[section]
\newtheorem{corol}{{\bf Corollary}}[section]
\newtheorem{propo}{{\bf Proposition}}[section]
\newtheorem{rem}{Remark}

\begin{document}

\title{Information Flow Rate for Cross-Correlated Stochastic Processes}

\author{Dionissios~T.~Hristopulos,~\IEEEmembership{Senior Member,~IEEE}
\thanks{D. Hristopulos is with the Department
of Electrical and Computer Engineering,  Technical University of Crete, Chania,
73100 Greece  (email: dchristopoulos@tuc.gr)}

\thanks{Manuscript received April XX, 2023; revised XXX XX, 2023.}}

\ifCLASSOPTIONpeerreview
\markboth{IEEE Transaction on Signal Processing,~Vol.~, No.~, XXXX~2023}%
{\MakeLowercase{Information Flow
Rate for Cross-Correlated Stochastic Processes}}
\else
\markboth{IEEE Transaction on Signal Processing,~Vol.~, No.~, XXXX~2023}%
{Hristopulos: \MakeLowercase{Information Flow
Rate for Cross-Correlated Stochastic Processes}}
\fi

\IEEEpubid{\makebox[\columnwidth]{978-1-5386-5541-2/18/\$31.00~\copyright2018 IEEE \hfill}
\hspace{\columnsep}\makebox[\columnwidth]{ }}

\ifCLASSOPTIONpeerreview
    \IEEEpeerreviewmaketitle
\else
    \maketitle
\fi

\IEEEpubidadjcol

\begin{abstract}
Causal inference seeks to identify cause-and-effect interactions  in coupled systems. A recently proposed method by Liang  detects causal relations by quantifying the direction and magnitude of information flow between time series.  The theoretical formulation of information flow for stochastic dynamical systems provides a general expression and a data-driven statistic for the rate of entropy transfer between different system units.
To advance understanding of information flow rate in terms of intuitive concepts and physically meaningful parameters, we investigate statistical properties of the data-driven information flow rate between coupled stochastic processes. We derive relations between the expectation of the information flow rate statistic and  properties of the auto- and cross-correlation functions. Thus, we elucidate the dependence of the information flow rate on the analytical properties and characteristic times of the correlation functions. 
Our analysis  provides insight into the influence of the sampling step, the strength of cross-correlations, and the temporal delay of correlations on information flow rate. We support the theoretical results with  numerical simulations of correlated Gaussian processes.
\end{abstract}


\IEEEpeerreviewmaketitle

\section{Introduction}
\label{sec:intro}

\IEEEPARstart{T}{here} is great interest in  the inference of causal relations between variables based on time series data.  Applications of causal inference extend across scientific disciplines such as neuroscience~\cite{Friston09,Hu11,Ning18}, bioinformatics~\cite{Ahmed20}, machine learning~\cite{Huang20,Bengio21,Nogueira22}, climate and Earth system sciences~\cite{Runge19,Kretschmer21,Diaz22}.
Classical measures of statistical association, such as the linear Pearson and the rank order (Spearman) correlation coefficients  fail to provide information about the directionality of interactions.  Hence, they can not reveal cause-effect relations. In fact, it  is well known that non-zero statistical correlations can be observed even in the absence of causal relations.   The latter are characterized by two key properties: (i) temporal precedence, i.e., the cause precedes the effect, and (ii) physical influence, i.e., variations of the cause have an impact on the  effect~\cite{Eichler13}.  Causal analysis investigates methods that can determine cause-effect relationships~\cite{Sommerlade09,He19,Hlavavckova07,Pearl09,Peters17,Smirnov18}.

Various methods of causal inference are in use. They include \emph{Wiener-Granger causality} (WGC)~\cite{Wiener56,Granger69,Siggiridou16,Siggiridou21} which is a standard tool for analyzing brain connectivity~\cite{Kaminski01,Hesse03,Bressler11,Marinazzo11}, kernel WGC~\cite{Marinazzo08,Marinazzo11} which generalizes WGC to nonlinear interactions, entropy-based methods~\cite{Hlavavckova07,Cohen14} such as transfer entropy (TE) and mutual information ~\cite{Schreiber00,Salvador10,Vicente11,Shovon14},  convergent cross mapping (CCM) which is based on the theory of dynamical systems~\cite{Sugihara12}, and PCMCI (Peter and Clark algorithm followed by momentary conditional independence test) which is based on Pearl’s graphical model framework and is applicable to nonlinear time series~\cite{Runge18,Runge19b}.

The \emph{Liang information flow rate} (IFR),  is  a non-parametric causality measure  between two potentially interacting time series~\cite{Liang08,Liang13,Liang14,Liang15}.
The  IFR  formulation  is based on Shannon entropy and the theory of dynamical systems. It leads to general expressions for information flow  between different variables which involve time-dependent expectations over  joint probability density functions~\cite{Liang08,Liang13,Liang13b,Liang14}.
The  IFR has been formulated---at least for two dimensional systems---using both absolute and relative entropy~\cite{Liang13,Liang14}. The relative entropy (Kullback–Leibler divergence)
quantifies the amount of information added to a system  with respect to the initial probability distribution.   IFR  was   recently extended to describe information flow in quantum mechanical systems~\cite{Yi22}.

In order to derive an estimate of IFR based on observations, Liang used a linear stochastic dynamical system that satisfies the following first-order stochastic differential  equation (also known as Langevin equation)
\beq
\label{eq:sode-motion-linear}
\di\bfX(t) = \mathbf{f}\, \di t + \mathbf{A}\,\bfX(t) \, \di t + \mathbf{B}\, \di\mathbf{W}(t) \, ,
\eeq
where $\di\bfX(t)$ is the differential of the two-dimensional stochastic process $\bfX(t)$,  $\mathbf{f}$ is a $2\times 1$ constant advection vector,  $\mathbf{A}$   is a $2\times 2$ matrix of (inverse) time constants, $\mathbf{B}$ is a $2\times 2$ diffusion matrix, and  $\di\mathbf{W}(t)$ is the differential of a  $2\times 1$ vector Wiener process (Brownian motion). Assuming that the observations represent a discretely sampled realization of $\bfX(t)$, he  derived
an  estimate of IFR  by maximizing the likelihood~\cite{Liang14}.
This data-driven  IFR  involves only the observed time series and their  finite differences. Hence, in contrast with transfer entropy which requires estimating bivariate probability distributions, and WGC which requires estimating potentially large and computationally intensive autoregressive systems---the complexity scales as $\Oo(N^{3}D^{3}p^{3})$, where $N$ is the sample size, $D$ is the number of components, and $p$ is the autoregressive order~\cite{Lindner19}---IFR has lower  computational complexity.

The data-driven IFR has been proven to recover causal relations in benchmark nonlinear systems~\cite{Liang18}. It has also been used to infer  brain connectivity from EEG  recordings~\cite{dth19} and fMRI data~\cite{Cong23}, to determine causal relations between global temperature and CO\textsubscript{2} concentration~\cite{Stips16}, to investigate interactions between climate modes (El Ni\~{n}o and the Indian Ocean Dipole)~\cite{Liang14}, to reconstruct sea surface height  using satellite altimetry~\cite{Rong22}, and to discover causal relations between the stocks of companies traded in the stock market~\cite{Liang15}.  Nonetheless, there are still gaps in our understanding of the data-driven IFR.

Machine learning methods such as Gaussian process regression, employ non-parametric models  that do not assume knowledge of dynamical equations. For  applications in which directional connectivity matters, it is important to select models that reflect the causal relations of the system. This necessity further motivates studying connections between IFR and stochastic processes.  In particular, the connection between different covariance kernel models and IFR properties has not been investigated.  Knowledge of such relations can guide the selection of suitable cross-covariance kernels for data-driven models based on Gaussian processes.  The dependence of IFR  estimates and their uncertainty on  statistical properties of the observed time series and different sampling steps are also important  for practical applications.   In addition,  the dependence of IFR on covariance kernel parameters can provide intuitive understanding of physical factors that influence IFR.

We seek to understand how  mathematical properties  (of the  covariance kernels) affect IFR.  Kernel properties can be learned from time series data using statistical estimation methods and optimal model selection approaches~\cite{Chen20}.  Our analysis  links the parameters and properties of the auto- and cross-covariance kernels with IFR.   The results are based on ensemble expectations that provide accurate IFR estimates in the ergodic limit.  We obtain an IFR expression  in terms of spectral moments, and we  investigate how  the continuity and differentiability of the stochastic processes  impact IFR. We also explore the dependence of IFR on the characteristic correlation times of the kernels and  the sampling step which  enables a deeper understanding of  information flow rate measurements.
Numerical simulations of cross-correlated Gaussian processes are also conducted to validate the theoretical results.

The  manuscript is organized as follows: Section~\ref{sec:prelim} presents notation and definitions. 
Section~\ref{sec:Cramer} focuses on permissibility conditions for the kernels of cross-correlated processes; these are necessary for constructing and simulating mathematically admissible  models with specified  properties.  In Section~\ref{sec:data-ifr} we discuss Liang's data-driven IFR and  its main mathematical properties.  Section~\ref{sec:ifr-ergodic} derives explicit equations for the IFR of cross-correlated, second-order ergodic processes (which satisfy the conditions laid out in Section~\ref{sec:Cramer}). In Section~\ref{sec:msd-processes} we focus on mean-square differentiable stochastic processes: we  obtain the continuous-sampling limit (which is applicable for very small sampling step) and leading-order corrections for finite time step.  Section~\ref{sec:msd-processes} investigates IFR properties  for mean-square-continuous (but non-differentiable) processes. These are suitable models for first-order stochastic   systems driven by Gaussian white noise.  Section~\ref{sec:simulations} presents numerical simulations which validate the main results of the theoretical analysis in the preceding sections.  Discussion related to IFR interpretation and  limitations of the analysis  is given in Section~\ref{sec:discussion}. Finally, Section~\ref{sec:conclusions} presents the main conclusions.

\subsection{Summary of main results}
The results obtained in this study for  IFR between coupled (cross-correlated) stochastic processes  are summarized below. 

\begin{enumerate}[wide, labelwidth=!, labelindent=0pt]
\item We define an expression for the equilibrium IFR which is valid  under ergodic conditions (Theorem~\ref{theor:IFR-ergodic}).  The equilibrium IFR allows connecting information flow with statistical properties of the processes involved that can be inferred from the data. We then calculate the equilibrium IFR in terms of spectral moments (Theorem~\ref{theor:IFR-ergodic-spectral}).  The spectral formulation provides existence conditions for  the equilibrium IFR that involve the spectral density tail.  

\item For small  sampling steps (compared to the characteristic correlation times), we obtain limit expressions of the  equilibrium IFR (Theorems~\ref{theorem:Tij-cs-msd}, \ref{theorem:Tij-cs-msd-delay}, \ref{theorem:Tij-cs-msc-delay}). These expressions depend on the regularity of the stochastic processes but are independent of the sampling step.   

\item We establish that the equilibrium IFR vanishes for the popular class of separable cross-correlation models (Proposition~\ref{propo:separ-zero-ifr}).  Hence, such models are not suitable for  causal analysis studies. 

\item  Our analysis proposes an interpretation of the IFR sign in the framework of stochastic processes.  The interpretation is related to the regularity of the processes (Theorems~\ref{theorem:Tij-cs-msd-delay}, \ref{theorem:Tij-cs-msc-delay}, \ref{theorem:ifr-ou-time-delay}). 

\item For processes with time-delayed correlations, we show that the IFR is a non-monotonic function of the time delay over the characteristic correlation time: it peaks for a  value of the ratio less than one and tends to zero as the ratio increases (Figs.~\ref{fig:gaussian-regression-simulated} and~\ref{fig:exponential-simulated}). 

\end{enumerate}

\section{Preliminaries}
\label{sec:prelim}

We use lowercase boldface symbols for vectors and uppercase boldfaced letters for matrices and vector stochastic processes (indexed by time).
The transpose of matrix $\bfA$ is denoted by $\bfA^\top$, its inverse by
$\bfA^{-1}$, and the  determinant by $\det\bfA$.  The sets of real and non-negative real numbers are denoted by $\R$ and  $\R_{\ge 0}$ respectively.
The set of natural numbers is denoted by $\Na$ and that of all integers by $\Z$.

\begin{defi}[Vector stochastic process]
A vector stochastic process with $D \in \Na$ components will be denoted by $\bfX(t) \triangleq \left[X_{1}(t), \ldots, X_{D}(t)\right]^\top$, where the symbol $\triangleq$ is used to  define a mathematical entity.  The stochastic process is defined on a probability space $(\Om, \mathcal{F}, P)$ and indexed by the time  $t \in \mathcal{T} \subset \R$, where $\mathcal{T}$ is an ordered set.
The  index $i=1, \ldots, D$ in $X_{i}(t)$ selects a specific scalar stochastic process.
\end{defi}

\smallskip

The \emph{expectation} of  $X_{i}(t)$ over the   ensemble space $\Omega$ is denoted by $m_{i}(t) \triangleq \E[X_{i}(t)]$. The \emph{fluctuation}  is denoted by $X'_{i}(t) \triangleq X_{i}(t) - \E[X_{i}(t)]$. 
Specific vector states (realizations) are denoted by $\bfx(t)$ and their components by $x_{i}(t)$, for $i=1, \ldots, D$.

\begin{defi}[Auto- and cross-covariance functions]
\label{defi:correl}
The covariance functions, $C_{i,j}(t, t+\tau)\triangleq \cov{X_{i}(t), X_{j}(t+\tau)}$, of two scalar processes $X_{i}(t)$ and $X_{j}(t)$ are defined by the  expectation (for $i,j=1, \ldots, D$)
\beq\label{eq:def-cov}
C_{i,j}(t, t+\tau) = \E[X'_{i}(t) \, X'_{j}(t+\tau)],
\eeq
where $\tau$ is the temporal lag, for all $t \in T$. For $i=j$ the \emph{auto-covariance} is obtained, while  for $i \neq j$~\eqref{eq:def-cov}  defines the \emph{cross-covariance}.  $\sigma^{2}_{i}= C_{i,i}(0)$, also denoted as $\var{X_{i}(t)}$, is the variance of the $i$-th component.
The \emph{auto-correlation functions (ACFs)} are then defined by  $\rho_{i,i}(\tau)= C_{i,i}(\tau)/ \sigma^{2}_{i}$ ($i=1, \ldots D$), while the \emph{cross-correlation functions (CCFs)} are given by $\rho_{i,j}(\tau)=C_{i,j}(\tau)/\sigma_{i} \sigma_{j}$, for $i \neq j=1, \ldots, D$.
\end{defi}

\medskip

\begin{defi}[Time series]
A stochastic vector process sampled at a discrete set of times will be denoted by $\{ \bfX_{t_n} \}_{n=1}^{N}$. Sampled values at $t_n$ will be denoted by $\bfx_{n}$ or $x_{i;n}$, where $n \in \Na$ is the time index. For a uniform time step $\delt$,  the sampling times are $t_{n} = n\delt$.
The set $\{ x_{i,n}\}_{n=1}^{N}$ , where $x_{i,n}\triangleq x_{i}(t_{n})$ denotes  a sample (time series) of the $i$-th stochastic process over $\{ t_{n} \}_{n=1}^{N}$.
\end{defi}

\smallskip

\begin{defi}[Stationarity]
A vector stochastic process $\bfX(t)$ is weakly or second-order stationary (henceforward, stationary) iff the following conditions hold: (i) $\E[\bfX(t)]=\bfm \in \R^{D}$ and (ii) $\E[X'_{i}(t-\tau)\,X'_{j}(t)]=C_{i,j}(\tau)$ for all $i,j=1, \ldots, D$. The minus sign preceding the time lag in $X_{i}(t-\tau)$  defines the CCF consistently with its use in the IFR formulation, see~\eqref{eq:t21-estim} below.
\end{defi}

\begin{defi}[Sample covariance functions]
We use the ``hat'' symbol for sample-based estimates of statistical quantities.
The sampling covariance function of  the time series $X_{i}(t), \, X_{j}(t)$ (where $i,j=1, \ldots, D$)  is given by
\beq
\label{eq:sample-cov}
\hC_{i,j}(k\delt) \triangleq \frac{1}{N-k} \sum_{n=k+1}^{N} x_{i,n-k}\, x_{j,n} - \overline{x_{i}}\, \overline{x_{j}}, \; \; k \in \Z \, ,
\eeq
where the ``overline''  denotes the sample average, i.e., $\overline{x_{1}}= \frac{1}{N} \sum_{n=1}^{N} x_{1,n}$. Equation~\eqref{eq:sample-cov} gives the sample auto-covariance for $i=j$ and the cross-covariance for $i \neq j$. 
The \emph{sampling correlation functions} are defined as follows
\beq
\label{eq:sample-cor}
\hrho_{i,j}(k\delt) \triangleq \frac{\hC_{i,j}(k\delt)}{\sqrt{\hC_{i,i}(0) \, \hC_{j,j}(0)}}.
\eeq

\end{defi}

\smallskip

\begin{defi}[Nonnegative definiteness]
A real, symmetric, $D \times D$ matrix $\mathbf{A}$ is nonnegative definite if
for any real-valued vector ${\mathbf z} \in \R^{D}$  it holds that $ {\mathbf z}^{\top} \mathbf{A} {\mathbf z} \ge 0$.  A function $C(t,t'): \, \R \times \R \to \R$ is  nonnegative definite iff
$\sum_{i=1}^{n} \sum_{j=1}^{n} z_{i} C(t_i, t_j) z_{j} \ge 0$ for all $n \in \Na$, all time vectors $(t_{1}, \ldots, t_{N}) \in \R^{n}$ and all vectors $(z_{1}, \ldots z_{n}) \in \R^n$.
\end{defi}

\begin{defi}[Fourier transforms]
The \emph{Fourier transform (FT)} of a  function $C(\tau): \R \to \R$ that is absolutely integrable  over the interval $(-\infty, \infty)$ is given by
\[
\widetilde{C}(\wm) = \int_{-\infty}^{\infty} \di\tau\, C(\tau)\, e^{-\imath \,\wm \tau}\,,
\]
where $\imath=\sqrt{-1}$ is the imaginary unit. 
In addition, if the function $C(\cdot)$ is of bounded variation in an interval which contains $\tau$, i.e., if $C(\cdot)$ has at most a finite number of extrema and discontinuities within this interval,   $C(\tau)$ is given by the inverse Fourier transform (IFT)~\cite{Priestley81}
\[
C(\tau) = \frac{1}{2\pi} \int_{-\infty}^{\infty} \di\wm \, \widetilde{C}(\wm)\, e^{\imath \wm \tau}.
\]

\end{defi}

A continuous function $C(\tau)$ is a permissible covariance kernel for some stationary scalar stochastic process if and only if it is nonnegative definite. Bochner's theorem~\cite{Bochner59} provides easily testable permissibility conditions.

\begin{theorem}[Bochner's permissibility theorem]
\label{theorem:bochner}
Let $C: R \to R$ be a continuous, absolutely integrable function. The function $C(\cdot)$ is non-negative definite if and only if its Fourier transform $\widetilde{C} \triangleq \mathrm{FT}[C]$ is nonnegative and integrable over $\R$.
\end{theorem}

\smallskip

The ergodic property allows replacing sample (temporal) with ensemble averages (i.e., expectations). It is useful in practical studies, because often only a single realization (sample) is available. We use ergodicity to investigate IFR. Slutsky's theorem provides necessary and sufficient conditions for second-order ergodic processes.   

\begin{theorem}[Slutsky's theorem]
\label{theorem:slutsky}
A second-order stationary, \emph{Gaussian} vector stochastic process  $\bfX(t)$ is second-order ergodic if and only if  $C_{i,j}(\tau) \to 0$ as $\tau \to \infty$ for all $i,j=1, \ldots, D$~\cite[pp.~526-533]{Papoulis02}.
\end{theorem}

\smallskip

\begin{rem}
\label{rem:slutsky}
For non-Gaussian processes, second-order ergodicity requires conditions on higher-order moments than the covariance.

\emph{Notation (Information flow):}
We denote information flow from the \emph{driver process} $X_{1}(t)$ to a \emph{receiver process} $X_{2}(t)$ by means of $X_{1}$\textrightarrow$X_{2}$.  $X_{1}$\textrightarrow$X_{2}$ implies that the equation which determines the dynamic evolution  of $X_{2}$ depends on $X_{1}$. In the context of~\eqref{eq:sode-motion-linear}, such dependence is expressed by means of $A_{2,1} \neq 0$ and/or $B_{2,1} \neq 0$.  Information flow  $X_{1}$\textrightarrow$X_{2}$  does not imply information flow in the reverse direction  $X_{2}$\textrightarrow$X_{1}$.

\section{Permissibility of Covariance Kernels for Multivariate Processes}
\label{sec:Cramer}

Bochner's theorem~\cite{Bochner59} applies  to scalar, second-order stationary processes. For vector stochastic processes, permissibility requires the stricter conditions  of Cram\'er's theorem~\cite{Cramer40}. This will be used below to construct valid separable and non-separable (time-delayed) covariance kernels  for cross-correlated processes. 

\medskip
\begin{theorem}[Cramer's Theorem]
\label{theor:cramer} The continuous matrix  function
$\bfC: \R \to \R^{D \times D}$
is a valid matrix covariance kernel for a continuous, stationary,  stochastic vector process, if the following conditions hold for the matrix components $C_{i,j}(\tau)$, for $i,j \in \{1, \ldots, D\}$:
\begin{enumerate}
\item[(C1)] The functions $C_{i,j}: \R \to \R$ are  absolutely integrable  (i.e., the FTs $\widetilde{C}_{i,j}$  exist) for all $i,j=1, \ldots, D$.
\item[(C2)]  The functions $C_{i,i}: \R \to \R$ satisfy Bochner's theorem  for all $i=1, \ldots, D$.

\item[(C3)]  The \emph{cross-spectral densities}   have bounded variation, i.e., the integrals
$\int_{\R} \di\wm \, \left|\widetilde{C}_{i,j}(\wm) \right|$ are finite for all $i \neq j =1,\ldots, D.$
\item[(C4)]  The spectral density matrix $\widetilde{\bfC}(\wm)$, where $[ \widetilde{\bfC}(\wm)]_{i,j}= \widetilde{C}_{i,j}(\wm)$,
is nonnegative definite for all  $\wm \in \R$.
\end{enumerate}

\end{theorem}

\medskip
The condition (C2)  establishes that $\tilde{C}_{i,i}(\wm)$
are permissible  \emph{auto-spectral densities} for scalar stochastic processes.
Establishing the permissibility of
$\tilde{C}_{i,j}(\wm)$  for $i \neq j$ is not trivial due to condition (C4) which requires that all eigenvalues or all the principal minors
of  $\widetilde{\bfC}(\wm)$  are nonnegative for all $\wm \in \R$.
To our knowledge, general methods for establishing the validity of (C4) are not available. The more restrictive concept of \emph{diagonal dominance} is often used to  derive sufficient conditions for matrix covariances~\cite{dth14}.

To circumvent the permissibility problem, the so-called separable (or intrinsic) model~\cite{Mardia93,Genton15} is often used.  It features a simple cross-correlation structure the permissibility of which is easily testable.

\smallskip
\begin{defi}[Separable cross-correlation model]
\label{defi:separable}
Let   $\mathbf{c}$ be a $D \times D$ positive-definite matrix with entries in $\R$ and $\rho(\tau): \R \to \R$  a non-negative definite function.  Then, the matrix function $\bfC(\tau)=\mathbf{c}\,\rho(\tau): \R \to \R^{D \times D}$ provides a permissible, separable  (\emph{intrinsic})  cross-correlation  model~\cite{Banerjee03}.
\end{defi}

\smallskip

While the separable model is demonstrably permissible, it is not very useful for studying information flow (see Proposition~\ref{propo:separ-zero-ifr} below.)  Hence, we introduce a more flexible model which involves time-delayed cross correlations.

\begin{lemma}[Time-delayed cross-correlations]
\label{lemma:time-delay-model}
(i) Let the continuous functions $C_{i,i}: \R \to \R$, $i=1,2$, be non-negative definite. (ii) Let   $C_{0}: \R \to \R$ be a continuous,  even  function, i.e., $C_{0}(\tau)= C_{0}(-\tau)$, of bounded spectral variation (cf. Condition C3 in Theorem~\ref{theor:cramer}) which has a global maximum at $\tau=0$. (iii) Define $C_{1,2}(\tau) \triangleq C_{0}(\tau-\tau_\ast)$, and $C_{2,1}(\tau) \triangleq C_{0}(\tau+\tau_\ast)$, where $\tau_\ast >0$ is the time delay.  (iv)  If the inequality  $D(\wm ) \triangleq \widetilde{C}_{1,1}(\wm)\widetilde{C}_{2,2}(\wm)- \widetilde{C}^{2}_{0}(\wm)>0$ holds for all $\wm \in \R$, the matrix function $\bfC(\tau)$ with  elements $C_{i,j}(\tau)$, $i,j \in \{1, 2\}$, is a valid matrix covariance function for the vector stochastic process ${\mathbf X}(t)=\left( X_{1}(t), X_{2}(t)\right)^\top$ which comprises  a \emph{leading (driver) series} $X_{1}$ and a \emph{lagging (receiver) series} $X_{2}$.
\end{lemma}

\begin{IEEEproof}
The conditions (C1)--(C3) of Cramer's theorem are satisfied by construction.
The definitions for $C_{1,2}(\tau)$, and $C_{2,1}(\tau)$ imply that the symmetry
$C_{1,2}(-\tau)=C_{2,1}(\tau)$ of the cross-covariance is satisfied since $C_{0}(-\tau-\tau_{\ast})= C_{0}(\tau+\tau_{\ast})$ due to the mirror symmetry of $C_{0}(\cdot)$.   The absolute integrability of $C_{0}(\tau)$ ensures the existence of the Fourier transform $\widetilde{C}_{0}(\wm)$. Furthermore, based on the time shift property of the Fourier transform it holds that
\begin{subequations}
 \begin{align}
\label{eq:cross-correl-delay}
\widetilde{C}_{1,2}(\wm)= e^{-\imath \wm \tau_\ast} \widetilde{C}_{0}(\wm), \\
\widetilde{C}_{2,1}(\wm)= \widetilde{C}_{1,2}^{\dag}(\wm)= e^{\imath \wm \tau_\ast} \widetilde{C}_{0}(\wm).
\end{align}
\end{subequations}

Therefore, $\lvert\widetilde{C}_{1,2}(\wm)\rvert= \lvert\widetilde{C}_{2,1}(\wm)\rvert=\lvert\widetilde{C}_{0}(\wm) \rvert$ and thus Condition (C3) is satisfied.
Finally,  it holds that $\widetilde{C}_{1,1}(\wm) \ge 0$, $\widetilde{C}_{2,2}(\wm) \ge 0$, and
$\widetilde{C}_{1,1}(\wm)\widetilde{C}_{2,2}(\wm) > \widetilde{C}^{2}_{0}(\wm)$ since $D(\wm)>0$ for all $\wm \in \R$, according to (iv) above. In light of~\eqref{eq:cross-correl-delay}, it also holds that $\widetilde{C}^{2}_{0}(\wm)=\widetilde{C}_{1,2}(\wm)\,\widetilde{C}_{2,1}(\wm)$.   Thus condition (C4) is satisfied for all $\wm \in \R$.
Furthermore, since $C_{1,2}(\tau) \triangleq \E[X_{1}(t-\tau)X_{2}(t)]$ attains its maximum value at $\tau = \tau_{\ast}$, the series $X_{1}$ leads and $X_{2}$ follows.
\end{IEEEproof}

\section{Data-Driven Information Flow Rate}
\label{sec:data-ifr}
Information flow  has been rigorously defined by means of an \emph{ab initio} approach~\cite{Liang16}. The latter involves calculating expectations over joint probability density functions that evolve dynamically. However, in many cases (e.g., earth systems science, neuroscience, mathematical finance) the only  information  comes from available data because the underlying stochastic dynamical system is not known \emph{a priori}. Liang developed a  data-driven IFR estimate by maximizing the likelihood of a  linear stochastic dynamical system with additive noise~\cite{Liang14}.  In the following,  we investigate this IFR statistic for systems of bivariate stochastic processes. For  systems involving $D>2$ processes, one can  consider all the pairwise combinations $C^{D}_{2}$.

For a pair of time series that represent realizations of two \emph{stationary stochastic processes} $X_{1}(t)$ and $X_{2}(t)$, the Liang IFR from  $X_{2}$ to  $X_{1}$, denoted by $2 \to 1$, is given by~\cite{Liang14}
\begin{subequations}
\label{eq:t21-estim}
\beq
\label{eq:t21-estim-1}
\htto(\delt) = \frac{\hr}{1 - \hr^{2}}\, \left[\,\hr_{2,d1}(\delt) - \hr\, \hr_{1,d1}(\delt) \,\right],
\eeq

\noi where $\hr \triangleq \hr_{1,2}$ is the linear (Pearson) correlation coefficient of  $\{x_{1,n}\}_{n=1}^{N}$ and $\{x_{2,n}\}_{n=1}^{N}$. Note that $\hr=\hat{\rho}_{1,2}(\tau=0)$ where  $\hat{\rho}_{1,2}(\tau)$ is the CCF defined in~\eqref{eq:sample-cor}.
The coefficients $\hr_{2,d1}(\delt)$ and $\hr_{1,d1}(\delt)$ represent correlations between the time series and their first-order finite differences,  defined by means of
\beq
\label{eq:t21-estim-2}
\hr_{i,dj}(\delt) \triangleq \frac{\hC_{i,dj}(\delt)}{\sqrt{\hC_{i,i}(0) \, \hC_{j,j}(0)}}, \; i, j=1, 2,
\eeq

\noi where $\hC_{i,dj}(\delt) \triangleq \E[X'_{i}(t)\,\dot{X}'_{j}(t)]$ is the sample covariance of the time series $\{ x_{i,n}\}$ and the  \emph{first-order difference} of $\{ x_{j,n}\}$, defined as
$\dot{x}_{j,n} \triangleq (x_{j,n+1} - x_{j,n})/\delt$, for $n=1, \ldots, N-1$.  Note that $\hr_{i,dj}(\delt)$ is not the standard correlation function between  $X'_{i}(t)$ and $\dot{X}'_{j}(t)$.
The correlation $\hr_{i,dj}(\delt)$ is equivalently expressed as
\begin{align}
\label{eq:hrij}
\hr_{i,dj}(\delt) = & \frac{\hC_{i,j}(\delt) - \hC_{i,j}(0)}{\,\delt\,\sqrt{\hC_{i,i}(0) \,\hC_{j,j}(0)}} =  \frac{\hrho_{i,j}(\delt) -\hrho_{i,j}(0) }{\delt},
\end{align}

\end{subequations}

\noindent where $\hC_{i,j}(\delt)$ and $\hrho_{i,j}(\delt)$ are respectively the sampling covariance  and correlation functions (auto- for $i=j$ and cross- for $i \neq j$) at lag  $\delt$ defined by means of~\eqref{eq:sample-cov} and~\eqref{eq:sample-cor}.

The general form of pair-wise IFR for a $D$-dimensional vector series $\bfX_{t}$ is given by (for $i,j =1, \ldots, D, \, i \neq j$)
\beq
\label{eq:tij}
 \htij(\delt)= \frac{\hr_{i,j}}{1 - \hr^{2}_{i,j}}\, \left[\,\hr_{i,dj}(\delt) - \hr_{i,j}\, \hr_{j,dj}(\delt)\, \right].
\eeq
Taking into account the cross-correlation expression~\eqref{eq:hrij} and the identity $\hrho_{j,j}(0)=1$, which is implied by~\eqref{eq:sample-cor}, we obtain the following equation for the IFR (for $i,j =1, \ldots, D, \, i \neq j$)
\beq
\label{eq:tij-no-ts-deriv}
\htij(\delt)= \frac{\hr_{i,j}}{1 - \hr^{2}_{i,j}}\, \frac{\,\hrho_{i,j}(\delt) - \hr_{i,j}\hrho_{j,j}(\delt)}{\delt}\,.
\eeq
In~\eqref{eq:tij-no-ts-deriv}, the $\hr_{i,dj}$ terms are replaced by expressions that involve only the sampling correlations $\hrho_{i,j}(\cdot)$.

\begin{rem}[General properties of $\htij$] The data-driven IFR satisfies the following general (independent of  the covariance kernels) properties (for $i, j=1, \ldots, D, \; i \neq j$):
\begin{itemize}
    \item For stationary processes, $\htij(\delt)$ is independent of the time index $t$.
    \item Since $\hr_{i,j}$ and $\hr_{i,dj}(\delt)$ are sample-based statistics, so are the $\htij(\delt)$.
    \item The definition~\eqref{eq:tij} implies that  $\htij(\delt)$ satisfies the following: (i) if $\hr_{i,j}=0$ then  $\htij=\htji=0$, and (ii) if $\htij \neq 0$, then $\htij(\delt) \propto 1/\delt$.
    \item The units of $\htij(\delt)$  are natural units of information (nats) per unit time.
	\item $\htij=0$ does not imply that necessarily $\htji=0$. 
\end{itemize}
\end{rem}

\section{Information Flow Rate for Second-Order Ergodic Processes}
\label{sec:ifr-ergodic}

For stationary processes, the data-driven IFR~\eqref{eq:tij} is  a  random variable that fluctuates between different system realizations.  We investigate the properties of $\htij(\delt)$ for cross-covariance-ergodic processes which satisfy  Theorem~\ref{theorem:slutsky}.

Practical use of Slutsky's theorem requires a sufficiently large sample to allow self-averaging. Hence, the length, $N\,\delt$, of the observation window  should be a large multiple of the longest correlation time among the functions $\{ C_{i,j}\}_{i,j=1}^{d}$. This condition can be expressed as $N \delt \gg \tau_{d}$, where
\[
\tau_{d} =  \max_{i, j=1, \ldots, d} \{\tau_{c;i,j} \}_{i, j=1}^{d}, \quad \tau_{c;i,j} = \int_{-\infty}^{\infty} C_{i,j}(\tau)\, \di\tau\, ,
\]
is the largest of the ACF, $\tau_{c;i,i}$  and CCF, $\tau_{c;i,j} (i \neq j)$ correlation times (measured in units of $\delt$).  This condition implies measuring the  time series length using an effective sample size (ESS) $\neff < N$, which accounts for correlation effects~\cite{Thiebaux84,Kass98}.  Different definitions of $\neff$~\cite{Bartlett46,Quenouille47,Afyouni19} agree on reducing $N$ by a factor that reflects the correlation times.  A typical estimate of the effective size is thus $\neff=N\, \delt/\tau_{d}$.
\end{rem}
\smallskip

\begin{theorem}[Equilibrium IFR for ergodic processes]
\label{theor:IFR-ergodic}
Let $\bfX(t)$ represent a  second-order  ergodic, $D$-vector, stochastic process with standard deviations $\sigma_i$, $i=1, \ldots, D$, auto- and cross-covariance functions $C_{i,j}(\tau)$, for $i,j=1, \ldots, D$, and   respective correlation functions $\rho_{i,j}(\tau)=C_{i,j}(\tau)/\sigma_{i} \sigma_{j}$.  

The \emph{equilibrium  IFR} between the process
$X_{i}$ and the process $X_{j}$, for $j \neq i$,  is given by  the $N \to \infty$ limit
\begin{align}
\label{eq:E-Tij}
T_{i\to j}(\delt)  & \triangleq \lim_{N \to \infty} \htij(\delt)   =
 \frac{\rho_{i,j}(0)}{1 - \rho_{i,j}^{2}(0)}\,
\nonumber \\
 & \quad \; \times \left[ r_{i,dj}(\delt) -  \rho_{i,j}(0)\, r_{j,dj}(\delt) \right],
\end{align}
\noi where   $r_{i,dj}(\delt)$ is the slope of the correlation function $\rho_{i,j}(\cdot)$ near zero lag, expressed as
\beq
\label{eq:ridj-ergodic}
r_{i,dj}(\delt) =  \frac{1}{\delt}\left[ \, \rho_{i,j}(\delt) - \rho_{i,j}(0) \, \right].
\eeq
\end{theorem}

\begin{IEEEproof}   Second-order ergodicity implies that the sample mean as well as the auto-covariance and cross-covariance functions converge in the mean square sense to their ensemble counterparts as $N \to \infty$.
If the ergodic conditions hold, the sample averages $\hr_{i,j}$ in~\eqref{eq:tij} can be replaced with their ensemble counterparts  $\rho_{i,j}(0)$.
For the $\hr_{i,dj}(\delt)$ terms, it holds that
\begin{align*}
\label{eq:ridia}
& \lim_{N \to \infty } \hr_{i,dj}(\delt) =   \frac{1}{ \sigma_{i}\, \sigma_{j}}\, \E\left[X'_{i}(t) \left( \frac{ X'_{j}(t+\delt) - X'_{j}(t)}{\delt} \right) \right]
\nonumber
\\[1ex]
 & \quad =   \frac{\left[ \, C_{i,j}(\delt) - C_{i,j}(0) \,\right]}{\delt\, \sigma_{i} \sigma_{j}} = \frac{ \rho_{i,j}(\delt) - \rho_{i,j}(0) \,}{\delt}.
\end{align*}
The last step uses Definition~\ref{defi:correl} for the correlation functions. The above leads to~\eqref{eq:ridj-ergodic} and  concludes the proof for~\eqref{eq:E-Tij}.
\end{IEEEproof}

\smallskip
Theorem~\ref{theor:IFR-ergodic} allows replacing the data-driven IFR in the ergodic limit with $\tij(\delt)$. The latter involves only ensemble moments (correlation functions) and can thus be used for theoretical   investigations of  information flow between stochastic processes. 

\smallskip
\begin{corol}[Equivalent expression for equilibrium IFR]
\label{corol:equilibrium-ifr}
If the conditions and definitions of Theorem~\ref{theor:IFR-ergodic} hold,  the equilibrium IFR between two second-order ergodic processes $X_{i}(t)$ and  $X_{j}(t)$ is given by
\begin{equation}
 \label{eq:equilibrium-ifr}
T_{i\to j}(\delt)= \frac{\rho_{i,j}(0)}{1 - \rho_{i,j}^{2}(0)}\, \frac{\,\rho_{i,j}(\delt) - \rho_{i,j}(0)\rho_{j,j}(\delt)}{\delt}.
 \end{equation}
 The above is expressed in terms of ACFs and CCFs of the two processes thus avoiding cross-correlations between the processes and their derivatives.
\end{corol}

\begin{IEEEproof}
The equation~\eqref{eq:equilibrium-ifr} is obtained from~\eqref{eq:E-Tij} in view of the correlation slope~\eqref{eq:ridj-ergodic} and the fact that $\rho_{j,j}(0)=1$.
\end{IEEEproof}

\medskip

Next, we show that for processes described by a separable cross-correlation model the equilibrium IFR vanishes. 

\smallskip

\begin{propo}[IFR for separable cross-correlation models]
\label{propo:separ-zero-ifr}
Consider  $D$ cross-correlated processes which  satisfy the covariance separability  condition of  Definition~\ref{defi:separable}. 
The equilibrium IFR $\tij(\delt)$ vanishes between any two non-fully correlated processes, i.e., $\tij(\delt)=0$ for $i\neq j \in \{1, \ldots, D\}$ if $\rho_{i,j}(0) \neq \pm 1$.
\end{propo}

\begin{IEEEproof}
Based on  Definition~\ref{defi:separable}, the correlation functions of separable models are given by
\begin{subequations}
\label{eq:separable-rho}
\begin{align}
\rho_{i,j}(\tau) = & \rho(\tau), \; \text{for} \; i=j,
\\
\rho_{i,j}(\tau) = & \frac{c_{i,j}}{\sqrt{c_{i,i}c_{j,j}}}\,\rho(\tau), \; \text{for} \; i \neq j\,.
\end{align}
\end{subequations}
The IFR satisfies $\tij(\delt) =A\,B(\delt)/\delt$, where $A \triangleq \rho_{i,j}(0)/(1 - \rho_{i,j}^{2}(0))$ and $B(\delt) \triangleq \rho_{i,j}(\delt) - \rho_{i,j}(0)\rho_{j,j}(\delt)$ according to Corollary~\ref{corol:equilibrium-ifr}. If $\rho_{i,j}(0) \neq \pm 1$, then $A \in \R$. If we define $a_{i,j} \triangleq \frac{c_{i,j}}{\sqrt{c_{i,i}c_{j,j}}}$,   it follows that $\rho_{i,j}(0)=a_{i,j}$, 
$\rho_{i,j}(\delt)=a_{i,j}\rho(\delt)$, and $\rho_{j,j}(\delt)=\rho(\delt)$, according to~\eqref{eq:separable-rho}; from the above  $B(\delt)=0$ is obtained.
\end{IEEEproof}

\smallskip 

Hence, there is no equilibrium information flow between stochastic processes with separable cross-correlation models. If the processes are fully correlated, i.e., $\rho_{i,j}(0)= \pm 1$, the IFR is not well-defined due to the zero term $1 - \rho_{i,j}^{2}(0)$ in the denominator of~\eqref{eq:equilibrium-ifr}.  In this case, however, the two processes differ by a sign at most; therefore, they are essentially the same process. 

Below we present a theorem which links the equilibrium IFR with the spectral moments of the CCFs.  As a consequence,  the equilibrium IFR is shown to vanish for $\delt \to 0$ if the CCF $\rho_{i,j}$ is an even function of $\tau$. 
\smallskip

\begin{theorem}[Spectral IFR expression for ergodic processes]
\label{theor:IFR-ergodic-spectral}
Let $\{X_{i}(t)\}_{i=1}^{D}$ represent a set of  stochastic processes with the properties specified in  Theorem~\ref{theor:IFR-ergodic}. Assume that $\widetilde{\rho}_{i,j}(\omega)$ are  the auto- and cross-spectral densities,  defined as the FTs of the correlation functions $\rho_{i,j}(\tau)$.  Furthermore, assume that
the auto- ($i=j$) and cross- ($i \neq j$) spectral moments of order one, $\Lambda_{i,j}^{(1)}$, defined by means of the improper integrals
\beq
\label{eq:spectral-moments}
\Lambda_{i,j}^{(1)} \triangleq
\frac{\imath}{2\pi}\int_{-\infty}^{\infty} \di\wm \, \wm \,   \widetilde{\rho}_{i,j}(\wm),
\eeq
exist.  Then, the equilibrium IFR is given by
\beq
\label{eq:E-Tij-spectral}
T_{i\to j}(\delt) =  \frac{\rho_{i,j}(0)}{1 - \rho_{i,j}^{2}(0)}\,  \Lambda_{i,j}^{(1)}  + \Oo(\delt), \;\; \text{for} \;\; i \neq j\,.
\eeq
\end{theorem}

\begin{IEEEproof}
Equation~\eqref{eq:E-Tij-spectral} follows from~\eqref{eq:E-Tij} by showing that $r_{i,dj}(\delt) \propto \Lambda_{i,j}^{(1)}+\Oo(\delt)$ for all $i,j$. We use~\eqref{eq:ridj-ergodic} for $r_{i,dj}(\delt)$, express $\rho_{i,j}(\cdot)$ in terms of its IFT, and replace  $\exp(\imath \wm t)$ with its Taylor expansion around $\delt=0$, to obtain
\begin{align}
\label{eq:ridj-spectral}
r_{i,dj}(\delt)   = & \frac{1}{2\pi \delt}\int_{-\infty}^{\infty} \widetilde{\rho}_{i,j}(\omega)\, \left(e^{\imath \wm \delt} - 1 \right)\, \di\wm
\nonumber \\
= & \frac{\imath}{2\pi}\int_{-\infty}^{\infty} \, \wm \,\widetilde{\rho}_{i,j}(\omega) \, \di\wm + \Oo(\delt)\,.
\end{align}

In general,  $\widetilde{\rho}_{i,j}(\omega)=\widetilde{\rho}_{i,j}^{\,\re}(\omega)+ \imath \,\widetilde{\rho}_{i,j}^{\,\im}(\omega)$.
Since $\rho_{i,j}(\delt)$ is a real-valued function,  its Fourier transform respects $\widetilde{\rho}_{i,j}^{\,\re}(\omega)=\widetilde{\rho}_{i,j}^{\,\re}(-\omega)$ and $\widetilde{\rho}_{i,j}^{\,\im}(-\omega)= - \widetilde{\rho}_{i,j}^{\,\im}(\omega)$. The integral involving $\omega \widetilde{\rho}_{i,j}^{\,\re}(\omega)$ over the symmetric interval $(-\infty, \infty)$ vanishes because $\widetilde{\rho}_{i,j}^{\,\re}(\omega)$ is  even, and thus the integrand $\omega \widetilde{\rho}_{i,j}^{\,\re}(\omega)$ is an odd function of $\omega$.  Hence, only the integral over the imaginary part survives in~\eqref{eq:ridj-spectral}, leading to
\beq
\label{eq:ridj-spectral-2}
r_{i,dj}(\delt) = - \frac{1}{\pi}\int_{0}^{\infty} \di\wm \, \wm \,   \widetilde{\rho}_{i,j}^{\,\im}(\wm)
= \Lambda_{i,j}^{(1)}.
\eeq
The calculation of  $r_{j,dj}(\delt)$ involves the density $\widetilde{\rho}_{j,j}^{\,\im}(\wm)$. The ACF ${\rho}_{j,j}(\tau)$ are real-valued, even functions. Hence,   the respective FTs are also real-valued and even. Thus,  $\widetilde{\rho}_{j,j}^{\,\im}(\wm)=0$ for all $\wm \in \R$, and the diagonal spectral moments $\Lambda^{(1)}_{j,j}$ vanish. Therefore, only the $r_{i,dj}(\delt)$ terms with $i \neq j$ give non-vanishing spectral integrals. Finally, the result~\eqref{eq:E-Tij-spectral} is obtained from~\eqref{eq:E-Tij} using the spectral integral~\eqref{eq:ridj-spectral-2}.
\end{IEEEproof}

\smallskip

\begin{rem}[Existence of spectral moments]
The existence of $\Lambda_{i,j}^{(1)}$  requires that the spectral densities $\widetilde{\rho}_{i,j}(\wm)$ satisfy $\widetilde{\rho}_{i,j}^{\,\im}(\wm) \sim \wm^{-2-\epsilon}$, where $\epsilon >0$ for $\wm \to \infty$.
The Whittle-Mat\'{e}rn spectral density is  $\widetilde{C}_{0}(\wm) \propto (1+\omega^{2} \lambda^2)^{-(\nu+1/2)}$, where $\nu>0$ is the smoothness index and $\lambda>0$ is a time constant~\cite{Genton15,dth20}. If $\nu=1/2$  the spectral integral~\eqref{eq:ridj-spectral-2} has a logarithmic divergence and $\Lambda_{i,j}^{(1)}$ is not well-defined. However, for $\nu>1/2$ the integral defining $\Lambda_{i,j}^{(1)}$ is finite.
\end{rem}

\begin{rem}[Vanishing of leading IFR term]
As it follows from~\eqref{eq:E-Tij-spectral} (Theorem~\ref{theor:IFR-ergodic-spectral}),   the leading IFR term vanishes if one the following conditions hold:
\begin{enumerate}[wide, labelwidth=!, labelindent=0pt]
\item If $\rho_{i,j}(0)=0$, i.e., if the processes are not cross-correlated at zero lag; then $T_{i\to j}(\delt)=T_{j\to i}(\delt)=0$.
\item If $\widetilde{\rho}_{i,j}^{\,\im}(\omega)=0$ for all $\omega >0$, i.e., if the FT of the  CCF is real-valued.  This leads to $\Lambda_{i,j}^{(1)}=0$ according to~\eqref{eq:ridj-spectral-2}.  A sufficient (but not necessary condition) is that $\rho_{i,j}(\tau)$ be an even function of $\tau$; then $\widetilde{\rho}_{i,j}^{\,\im}(\omega)=0$  for all $\omega \in \R$. 

In general,  $\widetilde{\rho}_{i,j}^{\,\im}(\omega)=0$ for $\omega>0$ does not imply $\widetilde{\rho}_{j,i}^{\,\im}(\omega)=0$ for $\omega>0$ since $\widetilde{\rho}_{j,i}^{\,\im}(\omega)= \widetilde{\rho}_{i,j}^{\,\im}(-\omega)$. So, if the leading-order term of $T_{i\to j}(\delt)$ vanishes, this does not imply that the leading-order term of $T_{j\to i}(\delt)$ also vanishes.
\end{enumerate}
\end{rem}

\section{IFR for Mean-Square Differentiable Processes}
\label{sec:msd-processes}

In the following we assume that the stochastic processes $\{X_{i}(t)\}_{i=1}^{D}$ are second-order  ergodic, and mean-square differentiable. For example, the displacement of a damped, linear harmonic oscillator driven by white noise is a mean-square differentiable process~\cite{dth23}. We will need the following lemma~\cite{Adler81,Cramer04,Papoulis02}.
\begin{lemma}[Mean-square continuity and differentiability]
\label{lemma:msd}
A second-order stationary process $X_{i}(t)$ is  continuous in the mean-square sense if its ACF $\rho_{i,i}(\tau)$ is continuous at $\tau=0$. $X_{i}(t)$ is  first-order differentiable (in the mean-square sense) if  $\rho_{i,i}(\tau)$ admits  a finite second-order derivative with respect to $\tau$ at zero lag, i.e., if $\rho_{i,i}^{(2)}(0)$ exists.  This condition implies that the first derivative of $\rho_{i,i}(\tau)$ vanishes at  $\tau=0$ (extremum condition).
\end{lemma}

\smallskip
We will consider the continuous-sampling limit $\delt \to 0$ where the sampling step is considerably smaller than  the shortest correlation time and the  delay $\tau_\ast$ (provided there is a finite delay $\tau_\ast$ between processes).  At this limit the leading term of $\tij(\delt)$  is dominant; this term is independent of $\delt$ according to~\eqref{eq:E-Tij-spectral} in Theorem~\ref{theor:IFR-ergodic-spectral}. Hence, at this limit the IFR has a value which is independent of $\delt$.
\medskip

\begin{theorem}[IFR continuous sampling limit]
\label{theorem:Tij-cs-msd}
Let $\{X_{i}(t)\}_{i=1}^{D}$ be a set of second-order  ergodic, and mean-square differentiable stochastic processes.
In addition, assume that the CCFs $\rho_{i,j}(\tau): \R \to \R$ are  at least twice differentiable (for $i \neq j$).  Let the  \emph{continuous-sampling limit of the IFR} be defined as follows:
\beq
\label{eq:tij-cs-lim}
\tijz \triangleq \lim_{\delt \to 0} \tij(\delt).
\eeq

\begin{enumerate}[wide, labelwidth=!, labelindent=0pt]
\item  $\tijz$ is a finite real number given by
\beq
\label{eq:tij-differentiable}
\tijz   = \frac{\rho_{i,j}(0) \, \rho_{i,j}^{(1)}(0)}{1 - \rho_{i,j}^{2}(0)},
\eeq
where $\rho_{i,j}^{(1)}(0)$ is the first-order derivative of the CCF  evaluated at zero lag. Furthermore, the \emph{sign} of $\tijz$ is the sign of the first-order derivative of $\rho_{i,j}^{2}(\tau)$ at $\tau=0$.

\item $\tijz$ has odd symmetry with respect to  interchange of the information flow indices, i.e., $\tijz= -\tjiz$.
\end{enumerate}
\end{theorem}

\begin{IEEEproof}
The proof is given in Appendix~\ref{app:Tij-differ}.
\end{IEEEproof}

\medskip

\begin{rem}[Connection with spectral formulation] The continuous-sampling IFR~\eqref{eq:tij-differentiable} is equivalent to the leading term in the spectral expression~\eqref{eq:E-Tij-spectral} since $\Lambda^{(1)}_{i,j}=\rho^{(1)}_{i,j}(0)$ based on~\eqref{eq:E-Tij-spectral} and $\mathrm{FT}[\rho_{i,j}^{(1)}]=\imath \omega\,\mathrm{FT}[\rho_{i,j}].$
\end{rem}

\smallskip

\begin{corol}[Finite-time-step corrections]
\label{corol:finite-step-msd}
If $\delt$ is small but not negligible, the equilibrium IFR is given by $\tij(\delt)=  \tijz +  \delta \tijz  +  \Oo(\delt^2)$ for $i \neq j$.  The first-order correction, $\delta\tijz$, to the continuous sampling IFR $\tijz$ is an $\Oo(\delt)$ term  given by
\begin{equation}
\label{eq:dtij-msd}
    \delta \tijz = \left\{ \begin{array}{ll}
       \tijz\,\left[\frac{\rho_{i,j}^{(2)}(0) -\rho_{i,j}(0)\,\rho_{j,j}^{(2)}(0)}{2\rho_{i,j}^{(1)}(0)} \right]\,\delt,    \text{if} \, \rho_{i,j}^{(1)}(0) \neq 0,
       \\[2ex]
       \frac{\rho_{i,j}(0)\,\left[ \rho_{i,j}^{(2)}(0) - \rho_{i,j}(0) \rho_{j,j}^{(2)}(0)  \right]\delt}{2\left[ 1- \rho_{i,j}^{2}(0)\right]},   \text{if} \, \rho_{i,j}^{(1)}(0) = 0\,.
    \end{array}  \right.
\end{equation}
\end{corol}

\begin{IEEEproof}
The proof is given in Appendix~\ref{app:finite-step-msd}.
\end{IEEEproof}

\smallskip

Above we provide general IFR expressions for mean-square differentiable processes.  Next, we focus on models with time-delayed correlations.  

\subsection{Cross-correlation functions with time delay}
\label{ssec:msd-delays}
Below we show that time-delayed cross-correlation models described in  Lemma~\ref{lemma:time-delay-model}  can sustain non-zero information flow $\tijz$, in contrast with separable models (cf. Proposition~\ref{propo:separ-zero-ifr}).

\medskip

\begin{theorem}[Continuous-sampling-limit IFR for  time-delayed cross correlations]
\label{theorem:Tij-cs-msd-delay}
Let $\bfC(\tau): \R \to \R^{2} \times \R^{2}$ represent a matrix covariance function which satisfies the conditions of Lemma~\ref{lemma:time-delay-model}.  Assume that at least the second-order derivatives of $\rho_{i,i}(\tau)$, $i=1,2$, and $C_{0}(\tau)$  with respect to $\tau$ exist at $\tau=0$. Then, the following statements are true:

\begin{enumerate}[wide, labelwidth=!, labelindent=0pt]
\itemsep0.5em
\item The continuous-sampling IFR is given by
\beq
\label{eq:tijz-msd-cs-td}
\tijz = 
\frac{C_{0}(-\epsilon_{i,j} \tau_{\ast}) \, C_{0}^{(1)}(u)\big\rvert_{u=-\epsilon_{i,j}\tau_{\ast}}}{\sigma^{2}_{i}\sigma^{2}_{j}-C_{0}^{2}(-\epsilon_{i,j} \tau_{\ast})},
\eeq
where $\sigma^{2}_{i}=C_{i,i}(0) >0$, $C_{0}^{(1)}(u)\rvert_{u=-\epsilon_{i,j}\tau_{\ast}}$ is the first derivative of $C_{0}(u)$  at $u=-\epsilon_{i,j}\tau_{\ast}$ and  $\epsilon_{i,j}$ is the \emph{Levi-Civita symbol}: $\epsilon_{i,j}=1$ if $(i,j)$ is an even  and $\epsilon_{i,j}=-1$ if $(i,j)$ is an odd permutation (e.g., $\epsilon_{1,2} =1; \epsilon_{2,1} =-1; \epsilon_{i,i} =0$).

\item $\tijz$ is antisymmetric, i.e.,
$\mathcal{T}_{i \to j} = - \mathcal{T}_{j \to i}$.
\end{enumerate}
\end{theorem}

\smallskip

\begin{IEEEproof}
The Levi-Civita symbol determines the sign of the time delay, i.e., $C_{i,j}(\tau)=C_{0}(\tau-\epsilon_{i,j}\tau_\ast)$.  We use~\eqref{eq:tij-differentiable}  from Theorem~\ref{theorem:Tij-cs-msd} for $\tijz$.  The value of $\rho_{i,j}(0)$ follows from $\rho_{i,j}(0)=C_{0}(-\epsilon_{i,j}\tau_{\ast})/\sigma_{i}\sigma_{j}$. Using the change of variable $\tau \mapsto u \triangleq \tau - \epsilon_{i,j}\tau_{\ast}$, it holds that
$\rho^{(1)}_{i,j}(0)= \alpha \frac{dC_{0}(\tau)}{d \tau} \rvert_{\tau=0}= \alpha \frac{dC_{0}(u)}{d u}\rvert_{u=-\epsilon_{i,j}\tau_{\ast}}$, where  $\alpha = 1/\left(\sigma_{i}\sigma_{j}\right)$. Then, we obtain
\beq
\label{eq:rij1-0}
\rho^{(1)}_{i,j}(0)= \frac{1}{\sigma_{i}\sigma_{j}}\left. \frac{d C_{0}(u )}{d u} \right|_{u=-\epsilon_{i,j}\tau_{\ast}},
\eeq
which concludes the proof of~\eqref{eq:tijz-msd-cs-td}.
The antisymmetry of $\tijz$ follows directly from (2) in Theorem~\ref{theorem:Tij-cs-msd}.
\end{IEEEproof}

\subsection{Square exponential covariance  with time delays}
\label{ssec:msd-gaussian-delayed}
We consider a bivariate stochastic process with square exponential auto-covariance functions  $C_{i,i}(\tau)=\sigma^{2}_{i}\, \exp(-\tau^{2}/\tau_{i}^{2})$ and  time-delayed cross-covariances generated  from the square exponential $C_{0}(\tau)= \sigma^{2}_{0}\, \exp(-\tau^{2}/\tau_{0}^2)$ as described in Lemma~\ref{lemma:time-delay-model}. The respective spectral densities are given by
$\widetilde{C}_{i}(\wm)= \sqrt{\pi}\,\tau_{i} \,\sigma^{2}_{i}\, \exp(-\wm^{2}\,\tau_{i}^{2}/4)$, for $i=0, 1, 2$.
The permissibility condition (C4) of Theorem~\ref{theor:cramer} requires that  $D(\wm) \ge 0$ for all $\wm \ge 0$, where
\begin{align}
\label{eq:gauss-condition}
D(\wm) \triangleq & \, \widetilde{C}_{1,1}(\wm)\, \widetilde{C}_{2,2}(\wm) - \widetilde{C}^{2}_{0}(\wm)
 \\
= & \pi \, \left[ \sigma_{1}^{2} \, \sigma_{2}^{2} \tau_{1}\, \tau_{2} \exo^{-\wm^{2}\left( \tau_{1}^{2} + \tau_{2}^{2} \right)/4}  - \sigma_{0}^{4} \, \tau_{0}^{2} \exo^{-\wm^{2} \tau_{0}^{2}/2} \right].
\nonumber
\end{align}

\noi The inequality $D(\wm) \ge 0$  is true for all $\wm \in \R$ provided that
(i) $\sigma_{1}^{2} \sigma_{2}^{2} \tau_{1} \tau_{2} \ge  \sigma_{0}^{4}\tau_{0}^{2}$ and (ii) $\tau_{1}^{2} + \tau_{2}^{2} \le 2 \tau_{0}^{2}$. These conditions ensure that $D(\wm=0)>0$ and that $\widetilde{C}_{1,1}(\wm)\, \widetilde{C}_{2,2}(\wm)$ decays more slowly than $\widetilde{C}^{2}_{0}(\wm)$.
The CCFs $(i \neq j)$ are expressed as  $\rho_{i,j}(\tau)=C_{0}(\tau -\epsilon_{i,j} \tau_{\ast})/\sigma_{1}\sigma_{2}$. Then,  $\rho_{i,j}(0)$ and  $\rho_{i,j}^{(1)}(0)$ are  given according to~\eqref{eq:rij1-0} by
\begin{align}
\label{eq:cross-gauss-delayed}
\rho_{1,2}(0) =   & \rho_{2,1}(0) = \frac{\sigma^{2}_{0}}{\sigma_{1}\sigma_{2}} \,\exo^{-\tau_{\ast}^{2}/\tau^{2}_{0}},
\\
\rho_{i,j}^{(1)}(0) = & \epsilon_{i,j} \frac{\sigma^{2}_{0}}{\sigma_{1}\sigma_{2}}\frac{2\tau_{\ast}}{\tau^{2}_{0}}\, \exo^{-\tau_{\ast}^{2}/\tau^{2}_{0}},  \; (i,j) =(1, 2), (2,1).
\end{align}
Based on the above and~\eqref{eq:tij-differentiable} we obtain the continuous-sampling IFR limit for $(i,j) \in \{(1,2), (2,1)\}$:
\beq
\label{eq:tij-time-delay-gauss}
\tijz = \frac{\rho_{i,j}(0) \, \rho_{i,j}^{(1)}(0)}{1 - \rho_{i,j}^{2}(0)} =
 \frac{2\tau_{\ast}\epsilon_{i,j}}{\tau_{0}^{2}}\frac{\sigma^{4}_{0}}{\sigma_{i}^{2}\sigma_{j}^{2} \exo^{2\tau_{\ast}^{2}/\tau^{2}_{0}} - \sigma^{4}_{0}}\,.
\eeq

\smallskip

\begin{rem}[IFR properties of mean-square-differentiable processes]
(1) The expression~\eqref{eq:tij-time-delay-gauss} for  ${\mathcal T}_{i \to j}$ does not depend on the characteristic ACF times $\tau_{1}$ and $\tau_{2}$ because the continuous-time equilibrium IFR~\eqref{eq:tij-differentiable} does not include diagonal terms. (2) The expression~\eqref{eq:tij-time-delay-gauss} implies, in light of $\sigma_{1}\sigma_{2}> \sigma_{0}^{2}$  (i) positive information flow rate from the leading (driver) time series, $X_{1}$ to the lagging (receiver) time series, $X_{2}$, and (ii) an equal-magnitude but opposite sign IFR   in the reverse direction  $X_{2} \to X_{1}$.  Furthermore, if we define the dimensionless variables $u \triangleq 2\tau_{\ast}^{2}/\tau_{0}^{2}$ and $w \triangleq \sigma_{1}^{2}\sigma_{2}^{2} / \sigma_{0}^{4}$, it  follows that
$\tijz = \epsilon_{i,j}u / \tau_{\ast}\left( w\,e^{u}-1\right)$. Hence,   $\tijz$ (and $\tjiz$ respectively) depends just on three parameters, the dimensionless ratios $u, w$ and the delay time $\tau_{\ast}$.
\end{rem}

\subsection{Linear regression model with delay}
\label{ssec:msd-regression-delayed}
As a special case of the bivariate stochastic process in Section~\ref{ssec:msd-gaussian-delayed}, we consider the  regression model
$X_{2}(t)= aX_{1}(t - \tau_{\ast}) + Y(t)$, where $0<a<1$ is a coupling factor, and $\tau_{\ast}$ is the time delay between the receiver process $X_{2}(t)$ and the driver process $X_{1}(t)$.  The latter is assumed to be a zero-mean  stationary process with square exponential covariance and characteristic time $\tau_{1}$;  $Y(t)$ is a zero-mean, stationary, mean-square differentiable  process with variance $\sigma^{2}_{y}=b^{2}\sigma_{1}^2$, where $\cov{X_{1}(t),Y(t')}=0$ for all $t,t' \in \R$ and $b$ is the relative (compared to $X_{1}$) amplitude  of the uncorrelated component $Y$ in the receiver.
Based on the above definitions, the variance of $X_{2}$ is  $\sigma^{2}_{2} = a^{2}\sigma^{2}_{1}+ \sigma^{2}_{y}$, while $\sigma^{2}_{0} =a\sigma^{2}_{1}$ is the cross-covariance at zero lag. In addition, the CCF is given by
\beq
\label{eq:rho12-linear-regression}
\rho_{1,2}(\tau)\triangleq \frac{\E[X_{1}(t-\tau)\, X_{2}(t)]}{\sigma_{1}\sigma_{2}}= \frac{a\, \rho_{1,1}(\tau-\tau_{\ast})}{\sqrt{a^{2} + b^{2}}}.
\eeq

\noindent The permissibility condition (C4) of Cramer's Theorem~\ref{theor:cramer}, as specified in inequality~\eqref{eq:gauss-condition},  is satisfied because (i) $\sigma_{1} \sigma_{2} = \sigma_{1}^{2} \sqrt{a^{2} + b^{2}} > \sigma_{0}^{2} =a\sigma_{1}^{2}$ and (ii)
$\tau_{1}^{2} = \tau_{2}^{2}=\tau_{0}^2$. Then, the expression~\eqref{eq:tij-time-delay-gauss}  for the continuous-sampling, equilibrium IFR yields the following for $(i,j) \in \{(1,2), (2,1)\}$:
\beq
\label{eq:Tij_differ_delayed}
\tijz =
 \frac{2\tau_{\ast}\epsilon_{i,j}}{\tau_{0}^{2}}
 \frac{1}{ \left(1 + b^2/a^2 \right)\, \exo^{2\tau_{\ast}^{2}/\tau^{2}_{0}} - 1}.
\eeq

According to~\eqref{eq:Tij_differ_delayed}, the magnitude of IFR is reduced as $b/a$ increases. This reflects the decline of the cross-correlation  between $X_{1}$ and $X_{2}$ with increasing $b/a$, as it follows from~\eqref{eq:rho12-linear-regression}.  The equilibrium IFR~\eqref{eq:Tij_differ_delayed}  is independent of the ACF of the ``contaminating'' signal $Y(t)$. The dependence of~\eqref{eq:Tij_differ_delayed} is further studied in the numerical simulations of Section~\ref{ssec:simul-gauss-time-delays}.

\section{IFR for Mean-Square Continuous but Non-differentiable Processes}
\label{sec:msc-processes}

First-order dynamical systems driven by Gaussian white noise generate stochastic processes which are mean-square continuous but not differentiable~\cite[p.~39]{Sarkka19}. 
If $X(t)$ is such a  stochastic process, its ACF   is continuous but non-differentiable   at zero lag. A well-known example is  the exponential (Ornstein-Uhlenbeck)  model with $\rho(\tau)= \exp(-|\tau|/\tau_{0})$; the lack of the first derivative at $\tau=0$ is due to the change in the slope of $\lvert \tau \rvert$ from positive to negative.  The  ergodic expression~\eqref{eq:equilibrium-ifr} for the equilibrium IFR is still valid in this case; however, the Taylor expansions of $\rho_{i,j}(\delt)$ (for $i=j$ and $i \neq j$) used in Theorem~\ref{theorem:Tij-cs-msd} are not.
In this section we calculate IFR for correlation functions that are continuous but non-differentiable at the origin. 

\subsection{Cross-correlation model with time delay}
\label{ssec:msc-delays}
We assume a time-delayed cross-correlation model as defined in Lemma~\ref{lemma:time-delay-model}, with the additional constraint that the covariance functions are continuous but non-differentiable at the origin.

\medskip

\begin{theorem}[Mean-square continuous processes with time-delayed cross correlations]
\label{theorem:Tij-cs-msc-delay}
Let $\bfC(\tau): \R \to \R^{2} \times \R^{2}$ represent a matrix covariance function as defined in Lemma~\ref{lemma:time-delay-model}.  Assume that  the ACFs $\rho_{i,i}(\tau)$, $i=1,2$, and the CCF generating function $C_{0}(\tau)$  are continuous and everywhere differentiable except at $\tau=0$.  In addition,  assume that $C_{0}(\cdot)$ is an even function of bounded spectral variation as specified in condition (ii) of Lemma~\ref{lemma:time-delay-model}. Then, the following statements are true for $i \neq j$:

\begin{enumerate}[wide, labelwidth=!, labelindent=0pt]
\itemsep0.2em

\item The continuous-sampling $(\delt \to 0)$ IFR limit is given by
\begin{align}
\label{eq:tij-cs-lim-continuous-delay}
\tijz =  &
\frac{r \, }{1-r^{2}}\left [\rho_{i,j}^{(1)}(0_{+}) - r \,\rho_{j,j}^{(1)}(0_{+})\right] \,,
\end{align}
where  $r \triangleq\rho_{i,j}(0)$,     $\rho_{i,j}^{(1)}(0_{+}) \triangleq C_{0}^{(1)}(-\epsilon_{i,j}\tau_{\ast})/\sigma_{i}\sigma_{j}$, and  the Levi-Civita tensor $\epsilon_{i,j}$ is defined in Theorem~\ref{theorem:Tij-cs-msd-delay}.

\item In the continuous-sampling limit we obtain
\beq
\label{eq:antisymmetry}
\tijz + \mathcal{T}_{j \to i} =
\frac{-r^{2} \, \left[ \rho_{i,i}^{(1)}(0_{+}) + \rho_{j,j}^{(1)}(0_{+}) \right]}{1-r^{2}} \ge 0\,.
\eeq
Hence, the IFR  antisymmetry  $\tijz = - \tjiz$ of mean-square differentiable processes (see Theorem~\ref{theorem:Tij-cs-msd}) is broken. 

\item For driver$\rightarrow$receiver IFR ($T_{1\to 2}$),  if $\delt<\tau_\ast$ the expression~\eqref{eq:tij-cs-lim-continuous-delay} is the leading-order  approximation in $\delt$. For  receiver$\rightarrow$driver IFR ($T_{2\to 1}$), \eqref{eq:tij-cs-lim-continuous-delay} is valid for small $\delt$ regardless of $\tau_\ast$.

\item  If $\delt > \tau_{\ast}$ (slow sampling regime), the leading-order  approximation of the  driver$\rightarrow$receiver IFR ($T_{1\to 2}$)  is given by
\beq
\label{eq:ifr-msc-small-delay}
T_{1\to 2}(\delt)= \frac{r\, \left[ \rho_{1,2}^{(1)}(0_{+}) \left(1 - \frac{2\tau_\ast}{\delt} \right) - r\, \rho_{2,2}^{(1)}(0_{+}) \right]}{1-r^{2}}\,.
\eeq
\end{enumerate}
\end{theorem}

\begin{IEEEproof}
The proof is given in Appendix~\ref{app:ms-continuous-delay}.
\end{IEEEproof}

\medskip

\begin{rem}[Zero IFR condition]
It follows directly from~\eqref{eq:tij-cs-lim-continuous-delay} that $\tijz$ vanishes  (i) if $r = 0$, i.e., if $\rho_{i,j}(0_{+})=0$, or (ii) if $\rho^{(1)}_{i,j}(0_{+})=r\, \rho_{j,j}^{(1)}(0_{+})$.
The  slope $\rho_{j,j}^{(1)}(0_{+})$ of the ACF at $\tau=0_{+}$ is negative. Hence, if $r>0$, condition (ii) is realized only if $\rho^{(1)}_{i,j}(0_{+})$, or equivalently   the first-order derivative of $C_{0}(\cdot)$, evaluated at $-\epsilon_{i,j}\tau_{\ast}$, is  negative (provided that $\tau_{\ast} >0$). Since $C_{0}(u)$ is a monotonically declining (increasing) function for $u>0$ ($u<0$), validity of condition (ii) requires $\epsilon_{i,j}<0$. Therefore, IFR can vanish   for $2 \to 1$ (lagging$\rightarrow$leading) information flow but not in the opposite, $1 \to 2$, direction. The situation is reversed if $r<0$; negative values for $r$ can only be obtained  if $C_{0}(\tau_\ast)<0$. 

\smallskip

A necessary (but not sufficient) condition for the separable correlation model  is $\tau_{\ast} = 0$.
The IFR $\mathcal{T}_{i\to j}$ in the limit $\tau_{\ast} \to 0$ is obtained from~\eqref{eq:tij-cs-lim-continuous-delay}  for flow in both directions.  Furthermore, in the separable case $\rho_{i,j}(\tau)=a_{i,j}\rho(\tau)$ (for $i \neq j$) while $\rho_{j,j}(\tau)=\rho(\tau)$ (cf. Proposition~\ref{propo:separ-zero-ifr}).  Hence, condition (ii) above is satisfied, and thus  $\tijz=\tjiz=0$  follows from~\eqref{eq:tij-cs-lim-continuous-delay}, in agreement with the more general result of Proposition~\ref{propo:separ-zero-ifr}.
\end{rem}

\smallskip

\subsection{Ornstein-Uhlenbeck (O-U) covariance model}
\label{ssec:msc-O-U}
In this section we study a concrete example of a bivariate process with exponential auto-covariance and time-delayed exponential cross-covariance. The exponential covariance  describes processes that obey the Ornstein-Uhlenbeck stochastic ordinary differential equation~\cite{OU30}.

\begin{corol}[Permissibility of O-U cross-covariance model]
\label{corol:permis-ou}
Consider the  exponential auto-covariance functions  $C_{i}(\tau)=\sigma^{2}_{i}\, \exp(-\lvert \tau/\tau_{i}\rvert)$, where $\tau_{i}>0$ for all $i \in \{1,2\}$, and the time-delayed cross-covariances generated from $C_{0}(\tau)= \sigma^{2}_{0}\, \exp(-\lvert \tau/\tau_{0} \rvert)$ as specified in Lemma~\ref{lemma:time-delay-model}.
Then, sufficient permissibility conditions are as follows:
\begin{subequations}
\begin{align}
\label{eq:O-U-permis}
\text{(C1).} & \quad \sigma_{1}\sigma_{2}\sqrt{\tau_{1} \tau_{2}} > \tau_{0} \sigma_{0}^{2},
\\[1ex]
   \text{(C2).} &  \quad \frac{\tau_{1}^{2} + \tau_{2}^{2}}{2\tau_{1} \tau_{2}} \ge \frac{\sigma_{1}^{2}\sigma_{2}^{2}}{\sigma_{0}^{4}} \ge \frac{\tau_{1} \tau_{2}}{\tau_{0}^{2}} \,.
\end{align}
\end{subequations}
\end{corol}

\begin{IEEEproof}
The respective  spectral densities for the O-U model are given by
$\widetilde{C}_{i}(\wm)= 2\tau_{i}\sigma^{2}_{i}/ (1+\wm^{2}\,\tau_{i}^{2})$, for $i=0, 1, 2$.
The permissibility conditions (C4) of Theorem~\ref{theor:cramer} require that
$\widetilde{C}_{1,1}(\wm) \ge 0$, $\widetilde{C}_{2,2}(\wm) \ge 0$, and
$D(\wm)>0$, for all $\wm \in \R$. The first two inequalities are valid for $\sigma_i, \tau_i >0$. The third inequality is expressed as
\[
\frac{\tau_{1}\tau_{2}\sigma_{1}^{2}\sigma_{2}^{2}}{(1+\wm^{2}\,\tau_{1}^{2})(1+\wm^{2}\,\tau_{2}^{2})} > \frac{\tau_{0}^{2} \sigma^{4}_{0}}{(1+\wm^{2}\,\tau_{0}^{2})^2} \equiv \frac{1}{D_{1}(\omega)} > \frac{1}{D_{2}(\omega)}\,,
\]
for all $\omega \in \R$, where $D_{i}(\wm)= \alpha_{i} + \beta_{i} \wm^{2} + \gamma_{i} \wm^{4}$ are quartic polynomials with coefficients
\begin{align*}
& \alpha_{1} = \frac{1}{\tau_{1}\tau_{2}\sigma_{1}^{2}\sigma_{2}^{2}}, \; 
\beta_{1}=\frac{\tau_{1}^{2} + \tau_{2}^{2}}{\tau_{1}\tau_{2}\sigma_{1}^{2}\sigma_{2}^{2}}, \; \gamma_{1}=\frac{\tau_{1}\tau_{2}}{\sigma_{1}^{2}\sigma_{2}^{2}}\,,
\\
& \alpha_{2}= \frac{1}{\tau_{0}^{2} \sigma^{4}_{0}}, \; \beta_{2}= \frac{2}{\sigma_{0}^{4}}, \; \gamma_{2}= \frac{\tau_{0}^{2}}{\sigma_{0}^{4}}
\end{align*}
Since $D_{i}(\wm) \ge 0$ for all $\wm \in \R$, the third inequality is equivalent to $D_{2}(\wm)>D_{1}(\wm)$ for all $\wm \in \R$. This condition is satisfied if $\alpha_2 > \alpha_1$, $\beta_2 \ge  \beta_1$, and $\gamma_2 \ge \gamma_1$, which lead to the conditions~\eqref{eq:O-U-permis} by simple algebraic calculations.
\end{IEEEproof}

\smallskip 
\begin{theorem}[IFR for O-U covariance model with time delay]
\label{theorem:ifr-ou-time-delay}
Assume that the O-U covariance model  satisfies the permissibility criteria of Corollary~\ref{corol:permis-ou}. Then, the \emph{continuous-sampling} IFR for the O-U model is given by
\beq
\label{eq:tij-time-delay-expon}
\tijz =
\frac{1 }{\left( \frac{\sigma^{2}_{1}\sigma^{2}_{2}}{\sigma^{4}_{0}}\right) \, \exo^{2\epsilon_{i,j} \tau_{\ast}/\tau_{0}} -1 }
\left(  \frac{1}{\tau_{j}} + \frac{\epsilon_{i,j}}{\tau_{0}}  \right).
\eeq
\end{theorem}

\begin{IEEEproof}
According to~\eqref{eq:rij1-0},  $\rho_{i,j}(0)=C_{0}(-\epsilon_{i,j}\tau_{\ast})/\sigma_{i}\sigma_{j}$ and the derivatives $\rho_{i,j}^{(1)}(0)$, for $i,j=1,2$,  are given by

\begin{subequations}
\label{eq:cross-o-u-delayed}
\begin{align}
\rho_{1,2}(0) =   &  \frac{\sigma^{2}_{0}}{\sigma_{1}\sigma_{2}} \exo^{-\tau_{\ast}/\tau_{0}}, \; \rho_{2,1}(0) = \frac{\sigma^{2}_{0}}{\sigma_{1}\sigma_{2}} \exo^{\tau_{\ast}/\tau_{0}}\,
\\
\rho^{(1)}_{i,j}(0_{+}) = & \frac{-\epsilon_{i,j}}{\tau_{0}}\, \left(\frac{\sigma^{2}_{0}}{\sigma_{1}\sigma_{2}} \right)\exo^{-\tau_{\ast}/\tau_0},  \; i \neq j,
\\
\rho_{j,j}^{(1)}(0_{+}) = & - \frac{1}{\tau_{j}}\,.
\end{align}
\end{subequations}
The first derivative $\rho^{(1)}_{i,j}(0_{+})$ (for $i\neq j$) is proportional to $\epsilon_{i,j}$. The Levi-Civita symbol controls the  slope which is negative for $(i,j)=(1,2)$ and positive for $(i,j)=(2,1)$.
If the permissibility criteria are satisfied,  \eqref{eq:tij-time-delay-expon} is obtained by plugging the expressions~\eqref{eq:cross-o-u-delayed} in the  IFR~\eqref{eq:tij-cs-lim-continuous-delay}, taking into account that $r=\rho_{i,j}(0)$.  
The expression~\eqref{eq:tij-time-delay-expon} is tested with numerical simulations in Section~\ref{ssec:simul-O-U-time-delay}.
\end{IEEEproof}
\smallskip

\begin{rem}[Asymmetric $\tijz$ for O-U model]
\label{rem:asymmetric-tij-msc}
If $\tau_0 = \tau_1$, then $\mathcal{T}_{2 \to 1}=0$ while $\mathcal{T}_{1 \to 2} \neq 0$.  This asymmetric behavior of IFR is in contrast with the square exponential model in which  both IFRs have the same magnitude and only differ in sign.
\end{rem}

Finally, according to~\eqref{eq:ifr-msc-small-delay}, for  $\delt > \tau_{\ast}$ (slow sampling) the leading-order with respect to $\delt$ leading$\rightarrow$lagging IFR  approximation is
\beq
\label{eq:ifr-expon-small-delay}
T_{1\to 2}(\delt)= \frac{2\tau_{\ast}}{\tau_{0}\,\delt}\,
\frac{1}{\left(\frac{\sigma_{1}^{2}\sigma_{2}^{2}}{\sigma_{0}^{4}} \right)
\exo^{2\tau_{\ast}/\tau_{0}} -1} \,.
\eeq

\section{Numerical Experiments}
\label{sec:simulations}
In this section we test the theoretical analysis of the equilibrium IFR
presented in Sections~\ref{sec:msd-processes}-\ref{sec:msc-processes} with synthetic time series generated by simulating zero-mean, Gaussian stochastic processes with specified correlation properties.
\subsection{Simulation Method}
\label{ssec:simulation-method}
We use the multivariate normal (MVN)  method
to simulate bivariate time series~\cite[p. 50]{Johnson87}. For time series comprising $N$  sampling times $t_{k} \in [0, T]$ for $k=1,\ldots, N$, the temporal dependence is determined by the full covariance
matrix  ${\bfC}$:
\beq
\label{eq:C-matrix}
{\bfC}= \left[\begin{matrix}
{\bfC}_{1,1} & {\bfC}_{1,2}  \\
 {\bfC}_{2,1} & {\bfC}_{2,2}
\end{matrix} \right]\,.
\eeq
$\bfC$ is a $2N \times 2N$ symmetric block matrix. The block submatrices $\bfC_{i,j}$, $i,j=1,2$, include
the $N \times N$  symmetric \emph{auto-covariance  matrices}  $\bfC_{1,1}, \bfC_{2,2}$, and the  $N \times N$ \emph{cross-covariance  matrices} $\bfC_{1,2}, \bfC_{2,1}$.  The matrix elements $[\bfC_{i,j}]_{k,l}=\mathbf{K}_{i,j}(t_{k}-t_{l})$ for $i,j=1,2$ and $k,l=1, \ldots, N$ are obtained from a $2 \times 2$ matrix function $\mathbf{K}(\tau)$ [e.g., see~\eqref{eq:cova-gaussian-simul} below].
If  ${\bfC} = {\bf A}{\bf A}^{T}$ is a factorization of ${\bf C}$,
 and  ${\bf z}$  is a $2N \times 1$ vector of  independent  random numbers drawn from the
 standard normal  distribution   $\mathcal{N}(0; 1)$,  then the $2N \times 1$ vector
 ${\bf x} = {\bf A}\,{\bf z}$  is a realization of the bivariate time series with said covariance structure. We use the
principal square root factorization of ${\bf C}$ for numerical stability~\cite{Higham87}.

\subsection{Square exponential covariance with time delay}
\label{ssec:gauss-time-delay-simul}

We simulate two  Gaussian processes governed by the delayed square exponential covariance model  (see Section~\ref{ssec:msd-gaussian-delayed}). We consider a  sampling window  of length $T=10$ and step  $\delt=0.002$, leading to $N=5\times 10^3$ sampling points.  
The covariance matrix function $\mathbf{K}$ is  given by
\beq
\label{eq:cova-gaussian-simul}
\mathbf{K}(\tau)= \left[\begin{matrix}
\sigma_{1}^{2}\, \exo^{-\tau^{2}/\tau_{1}^{2}} & \sigma_{0}^{2}\, \exo^{-(\tau - \tau_{\ast})^{2}/\tau_{0}^{2}}  \\[2ex]
 \sigma_{0}^{2}\, \exo^{-(\tau + \tau_{\ast})^{2}/\tau_{0}^{2}}  & \sigma_{2}^{2}\, \exo^{-\tau^{2}/\tau_{2}^{2}}
\end{matrix} \right]\, ,
\eeq
where
$\tau_{0}=T/200$,  $\tau_{1}=\tau_{2}=\tau_{0}$, $\tau_{\ast}=0.008$, $\sigma_{1}^{2}=\sigma_{2}^{2}=1.1$ and $\sigma_{0}^{2}=1$. For these parameters, the CCF at zero lag is $\rho_{1,2}(0) \approx 0.89$ according to~\eqref{eq:cross-gauss-delayed}. The small ratio $\tau_{0}/T$ is selected in order to ensure that $\neff \triangleq N\, \delt/\tau_{0}= T/\tau_{0} \gg 1$ in compliance with ergodic conditions (see Remark~\ref{rem:slutsky}).

\begin{figure}%
\centering
\subfloat{%
\label{fig:gauss-cova}%
\includegraphics[width=0.8\linewidth]{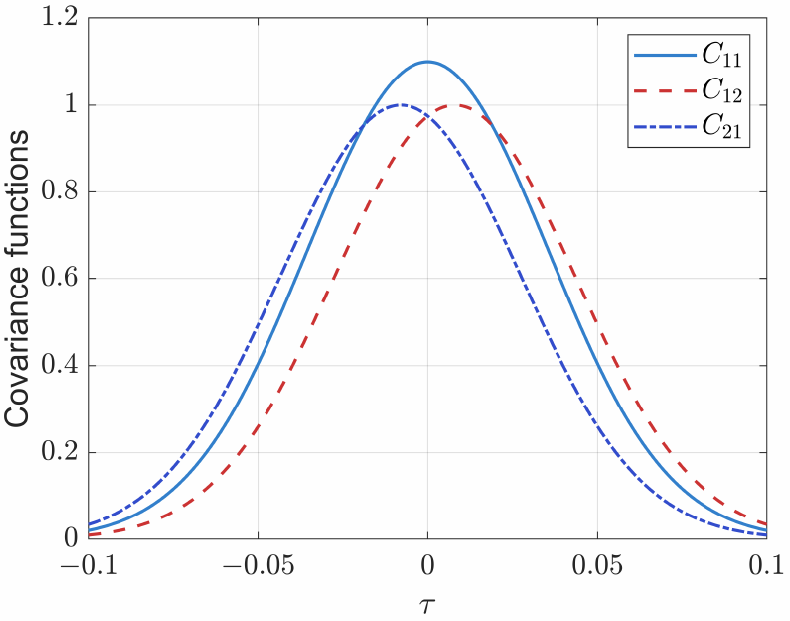}}
\\
\subfloat{%
\label{fig:bivariate-realizations}%
\includegraphics[width=0.8\linewidth]{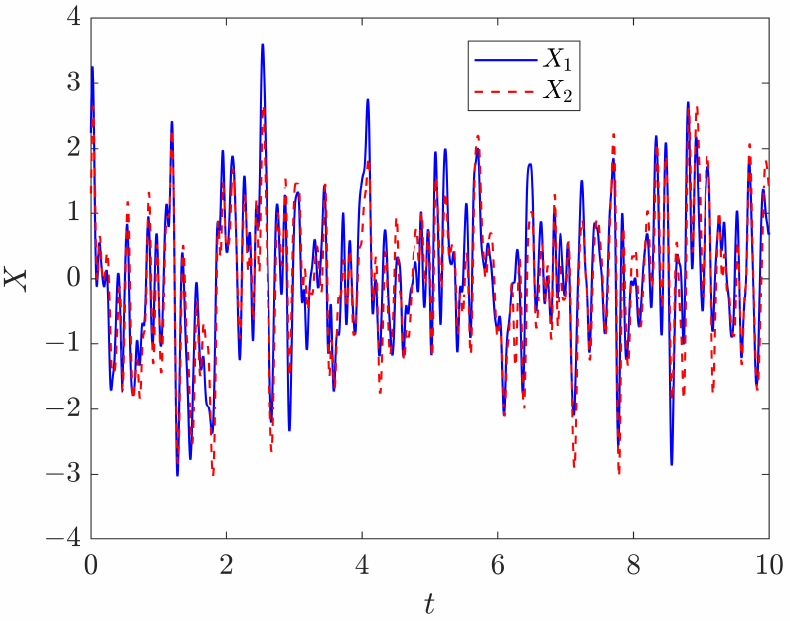}}
\caption[]{Auto- and cross-covariance functions (top) and simulated samples (bottom) of two Gaussian processes representing a driver time series $X_{1}(t)$ (continuous line, blue online) and  a receiver time series $X_{2}(t)$ (broken line, red online). The plots are based on the bivariate Gaussian process with the delayed square exponential covariance model~\eqref{eq:cova-gaussian-simul}.  In the top frame, the auto-covariance is marked by the continuous line (cyan online), the cross-covariance $C_{1,2}(\tau)$ by the broken line (red online), and the cross-covariance $C_{2,1}(\tau)$ by the dash-dot line (blue online).}%
\label{fig:bivariate-simul}%
\end{figure}

\begin{figure}
\centering
\includegraphics[width=0.8\linewidth]{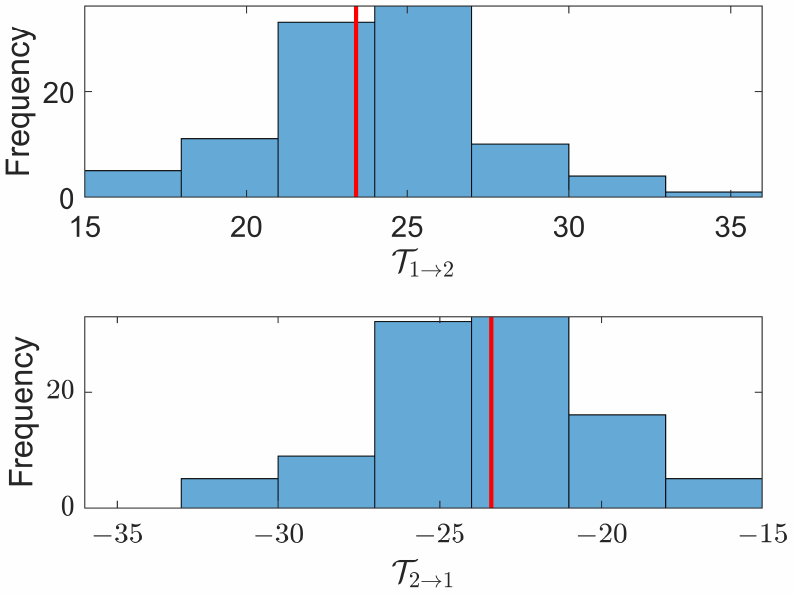}
\caption{Histograms of continuous sampling IFR $\mathcal{T}_{1 \to 2}$ (top) and $\mathcal{T}_{2 \to 1}$ (bottom) generated from an ensemble of 100 realizations of a bivariate cross-correlated Gaussian stochastic process with square exponential covariance model~\eqref{eq:cova-gaussian-simul}. The continuous line in the middle (red online) of both plots marks the theoretical estimate of the equilibrium IFR~\eqref{eq:tij-time-delay-gauss}.
The model parameters  for the stochastic processes (cf.~Section~\ref{ssec:gauss-time-delay-simul}), are:  $T=10$ (length of sampling window)   $\delt=0.002$ (sampling step),  $N=5\times 10^3$ (number of sampling points), $\tau_{1}=\tau_{2}=\tau_{0}=0.05$ (correlation time), $\tau_{\ast}=0.008$ (time delay), $\sigma_{1}^{2}=\sigma_{2}^{2}=1.1$ and $\sigma_{0}^{2}=1$.}
\label{fig:gaussian-delayed}
\end{figure}

An ensemble of $N_{\mathrm{sim}}=100$ realizations  is generated using MVN simulation (outlined in Section~\ref{ssec:simulation-method}).  The ACFs and CCFs along with a typical realization are shown  in Fig.~\ref{fig:bivariate-simul}.  The equilibrium IFR  is obtained from~\eqref{eq:tij-time-delay-gauss}.  The data-driven IFR is calculated from each realization using~\eqref{eq:tij}. The histograms of $\mT_{1 \to 2}$ and $\mT_{2 \to 1}$ calculated from the ensemble are shown in Fig.~\ref{fig:gaussian-delayed}.   The two histograms are nearly identical mirror images as expected from Theorem~\ref{theorem:Tij-cs-msd-delay}.  The theoretical values (marked by continuous vertical lines in Fig.~\ref{fig:gaussian-delayed}) lie in the middle of the respective histograms. According to~\eqref{eq:tij-time-delay-gauss}, $\mT_{1\to 2} \approx 23.4$ whereas the simulation-average yields  $\overline{\mT}_{1\to 2} \approx 23.75$.   In addition, the sample estimates of $\mT_{1\to 2}$ ($\mT_{2\to 1}$)  obtained from the simulated states are consistently positive (negative).

\paragraph{IFR dispersion}
The  IFR values exhibit broad distributions with a coefficient of variation $\approx 0.15$ (in absolute value). This behavior is due to between-samples fluctuations of  $\hr$, and the $1-\hr^{2}$ dependence of the IFR  [cf.~\eqref{eq:tij}].  The IFR variance thus tends to increase as $\hr \to \pm 1$.
In spite of the fluctuations,  the  IFR probability distributions preserve the  signs of $T_{1 \to 2}$ and $T_{2 \to 1}$ (see Fig.~\ref{fig:gaussian-delayed}), implying that the data-driven IFR correctly captures the direction of information flow from the leading to the lagging series.

\paragraph{Finite-sampling-step corrections} To obtain accurate IFR estimates  based on the continuous sampling expression~\eqref{eq:tij-time-delay-gauss}, it is necessary that the $\Oo(\delt)$ correction, $\delta \tijz$, be negligible. Based on~\eqref{eq:dtij-msd} the  magnitude of the relative correction is
\[
\left\lvert \frac{\delta \tijz}{\tijz} \right\rvert = \left\lvert \frac{\rho_{i,j}^{(2)}(0) -\rho_{i,j}(0)\,\rho_{j,j}^{(2)}(0)}{2\rho_{i,j}^{(1)}(0)} \right\rvert \,\delt\,.
\]
Since $\rho_{j,j}(\tau)=\exp(-\tau^{2}/\tau_{j}^{2})$ and  for $i \neq j$ it holds that  $\rho_{i,j}(\tau)=(\sigma_{0}^{2}/\sigma_{1}\sigma_{2})\,\exp\left[ -(\tau - \epsilon_{i,j}\tau_{\ast})^2\right]$,  using straightforward algebraic operations, it follows that
\beq
\label{eq:relative-correction}
\left\lvert \frac{\delta \tijz}{\tijz} \right\rvert = \frac{\delt}{2\tau_{\ast}} \left( \frac{\tau_{0}^{2}}{\tau_{j}^{2}} -1\right) + \frac{\delt\,\tau_{\ast}}{\tau_{0}^{2}}\,.
\eeq
The first term on the right-hand side of~\eqref{eq:relative-correction} vanishes since $\tau_0 = \tau_{1} = \tau_{2}$. Using the specified values of $\tau_0$, $\tau_\ast$ and $\delt$, we obtain
$\lvert \delta \tijz / \tijz \rvert \approx 6.4 \times 10^{-3}$ which is indeed negligible.

\subsection{Linear regression model with time delay}
\label{ssec:simul-gauss-time-delays}
We  test the  validity of the theoretical IFR~\eqref{eq:Tij_differ_delayed}  for the linear regression model of mean-square differentiable processes presented in Section~\ref{ssec:msd-regression-delayed}.
The study design is described below.

\begin{enumerate}[wide, labelwidth=!, labelindent=0pt]

\item The  processes $X_{1}(t)$ and $X_{2}(t)$ are coupled by means of the delayed square exponential covariance~\eqref{eq:cova-gaussian-simul}. The model parameters are  $\tau_{0}=T/200$,  $\tau_{1}=\tau_{2}=\tau_{0}=1$,  $\sigma_{1}^{2}=1.1$ and $\sigma_{0}^{2}=1$. Note that $a=\sigma^{2}_{0}/\sigma^{2}_{1}= (1.1)^{-1}$, while $\sigma^{2}_{2}=a^{2}\sigma^{2}_{1}\left(1+ b^{2}/a^{2} \right)$ varies with $b/a$.

\item The equilibrium IFR  is calculated for 100 values of  the ratio $\tau_{\ast}/\tau_{0} \in [0, 1]$ and for $b/a \in \{0.1, 0.2, 0.4, 0.6, 0.8\}$.

\item An ensemble of $N_{\textrm{sim}}=100$ simulations is generated by means of the MVN method for each    combination $(\tau_{\ast}/\tau_{0}, b/a)$ (see Section~\ref{ssec:simulation-method}).

\item Each pair of time series in the ensemble comprises $N=1000$ time instants $t \in [0, 100]$.

\item For each realization, we calculate $\mathcal{T}_{1\to2}$ and $\mathcal{T}_{2\to1}$ based on the data-driven IFR~\eqref{eq:tij}.

\item We generate parametric plots of the equilibrium IFRs versus $\tau_{\ast}/\tau_{0}$ for the different $b/a$ values.

\end{enumerate}

\begin{figure}
\centering
\includegraphics[width=0.95\linewidth]{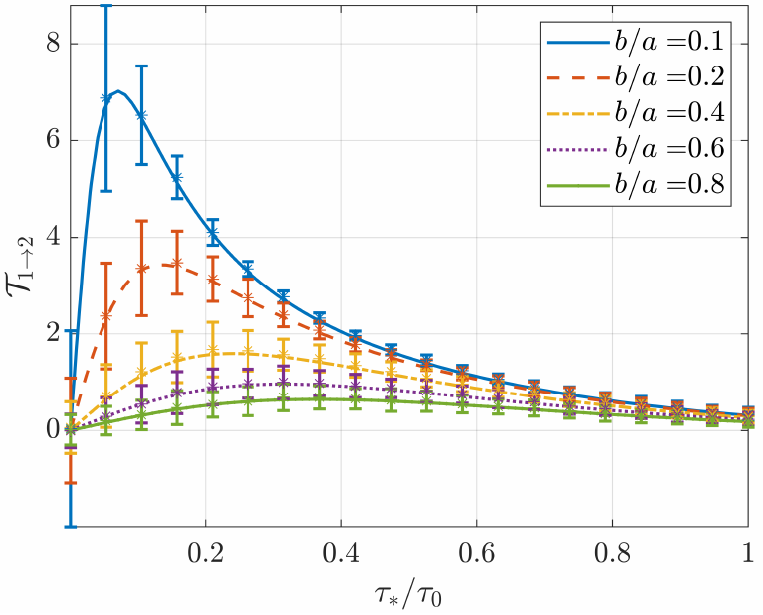}
\caption{Parametric plots of ${\mathcal T}_{1\to2}$ for the regression model $X_{2}(t)=a X_{1}(t-\tau_{\ast})+ Y(t)$ (see Section~\ref{ssec:msd-regression-delayed}) versus $\tau_{\ast}/\tau_{0}$ for different  $b/a$ ratios. $X_{1}(t), X_{2}(t)$ and $Y(t)$ are Gaussian processes with square exponential auto- and cross-covariances. It is assumed that $\sigma^{2}_{0}=1$, $\sigma^{2}_{1}=1.1$, $a=\sigma^{2}_{0}/\sigma^{2}_{1}$  and $\sigma_{2}^{2}= a^{2}\,\sigma_{1}^{2} (1 +b^{2}/a^{2})$. The time constants are $T=100$, $\tau_{1}=\tau_{2}=\tau_{0}=1$, while the time delay $\tau_{\ast}$ is determined from the ratio $\tau_{\ast}/\tau_{0}$. The continuous curves are based on the theoretical equilibrium IFR  expression~\eqref{eq:Tij_differ_delayed}. For every realization (pair of length $N=1000$  time series)  from an ensemble of $N_{\mathrm{sim}}=100$ simulations, the IFR is estimated based on  the data-driven estimator~\eqref{eq:tij}.  Lower values of $b/a$ imply higher cross correlation between $X_1$ and $X_2$.  The star markers denote  ensemble averages of IFR estimates, while the associated error bars  have a width of two standard deviations  (as estimated from the ensemble).
}
\label{fig:gaussian-regression-simulated}
\end{figure}

Figure~\ref{fig:gaussian-regression-simulated} compares the ensemble averages of the IFR $\mathcal{T}_{1\to2}$ with the  equilibrium IFR~\eqref{eq:Tij_differ_delayed}. The error bars are based on $2\sigma_{\textrm{IFR}}$, where $\sigma_{\textrm{IFR}}$ is the IFR standard deviation estimated from the simulation ensemble. Near-perfect  agreement is observed between the theoretical and the simulation-based estimates. The IFR peaks at a $\tau_{\ast}/\tau_{0}$ value which depends on $b/a$. The IFR (both theoretical and ensemble estimates) tend to zero as $\tau_{\ast}/\tau_0 \to 0$ in agreement with Proposition~\ref{propo:separ-zero-ifr}.  The dispersion of simulation-based IFR values  declines with increasing $\tau_{\ast}/\tau_{0}$ (for fixed $b/a$) and with increasing $b/a$ (for fixed $\tau_{\ast}/\tau_{0}$).  This behavior agrees with~\eqref{eq:rho12-linear-regression} which shows that $\rho_{1,2}(0)$ declines with increasing $b/a$ (under fixed $\tau_{\ast}$) and with increasing $\tau_\ast$ (under fixed $b/a$).  Values of $\rho_{1,2}(0)$ (and thus of $\hr$) that approach $\pm 1$ inflate the $\mathcal{T}_{1\to 2}$ variance. 
Finally, according to~\eqref{eq:relative-correction} the relative magnitude of second-order corrections to the theoretical estimate is $\sim \delt \,\tau_{\ast}/\tau_{0}^2$. Given the values of $\delt, \tau_{0}$ and the range of $\tau_{\ast}/\tau_0$, such corrections are  negligible. The results for $\mathcal{T}_{2\to1}$ are not shown since they are practically  mirror images (with reversed sign) of those for $\mathcal{T}_{1\to2}$.

\subsection{Ornstein-Uhlenbeck covariance with time delay}
\label{ssec:simul-O-U-time-delay}
We study two coupled, mean-square continuous Gaussian processes governed by an
exponential (O-U) covariance model with time delay. The process parameters $N,  \tau_{l}$ and $\sigma_{l}^{2}$, where $l=0,1,2$, the time step $\delt$, the observation window $T$ and the number $N_{\mathrm{sim}}$ of simulated states take the values used in  Section~\ref{ssec:msc-O-U}. The auto-covariance functions are given by  $C_{i,i}(\tau)=\sigma^{2}_{i} \, \exp(-\lvert \tau \rvert/\tau_{i})$, for $i=1,2$, and the cross-covariance functions are given by the exponentials 
$C_{i,j}(\tau)=\sigma^{2}_{0}\exp(-\lvert \tau -\epsilon_{i,j} \tau_{\ast}\rvert/\tau_0)$, for $(i,j)=(1,2), (2,1)$.

\begin{figure}[!h]
\centering
\includegraphics[width=0.8\linewidth]{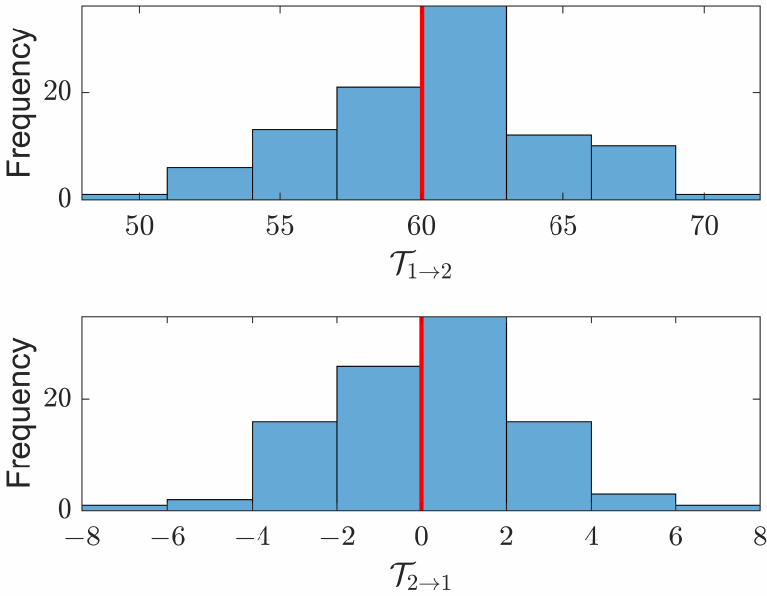}
\caption{Histograms of continuous sampling IFR $\mathcal{T}_{1 \to 2}$ (top) and $\mathcal{T}_{2 \to 1}$ (bottom) generated from an ensemble of 100 realizations comprising two Gaussian stochastic processes governed by the time-delayed, exponential  (Ornstein-Uhlenbeck)  model which is defined in Section~\ref{ssec:msc-O-U}. The vertical lines in the middle of the histograms (red online) near 60 and 0 respectively, mark the theoretical, equilibrium IFR estimate~\eqref{eq:tij-time-delay-expon}. The model parameters  for the stochastic processes are:  $T=10$ (length of sampling window)   $\delt=0.002$ (sampling step),  $N=5\times 10^3$ (number of sampling points), $\tau_{1}=\tau_{2}=\tau_{0}=0.05$ (correlation time), $\tau_{\ast}=0.008$ (time delay), $\sigma_{1}^{2}=\sigma_{2}^{2}=1.1$ and $\sigma_{0}^{2}=1$.}
\label{fig:expon-delayed}
\end{figure}

The IFR histograms  obtained from $100$ realizations are shown in Fig.~\ref{fig:expon-delayed}.  Both plots agree with the theoretical result~\eqref{eq:tij-time-delay-expon}. In addition, both $\mT_{1 \to 2}$ and $\mT_{2 \to 1}$  exhibit significant dispersion (due to the fluctuations of  $1/(1-\hr^2)$ as discussed in Section~\ref{ssec:gauss-time-delay-simul}).  However, $\mT_{1 \to 2}$ is persistently non-negative, demonstrating information flow from the leading to the lagging time series, while  $\mT_{2 \to 1}$ fluctuates around zero as expected based on~\eqref{eq:tij-time-delay-expon} and the fact that $\tau_{1}=\tau_{2}=\tau_{0}$.

\begin{figure}%
\centering
\subfloat{
\label{fig:exponential-simulated-1to2}%
\includegraphics[width=0.8\linewidth]{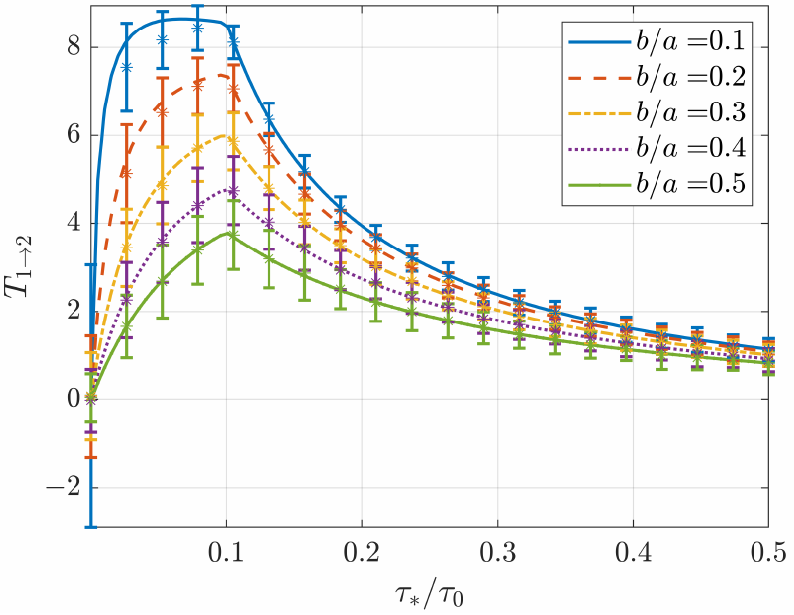}}
\\
\subfloat{
\label{fig:exponential-simulated-2to1}%
\includegraphics[width=0.8\linewidth]{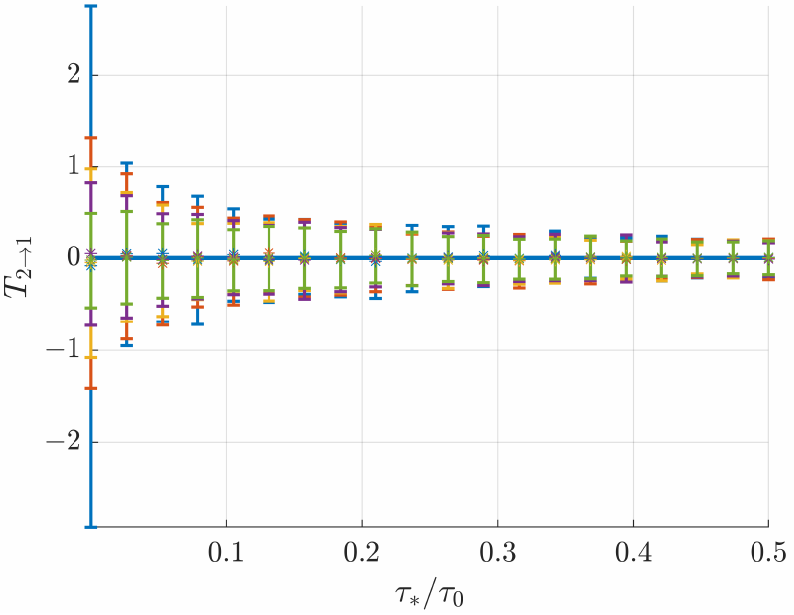}}
\caption[]{Parametric plots (continuous curves) of $T_{1\to2}$ (top) and $T_{2\to1}$ (bottom) versus $\tau_{\ast}/\tau_{0}$ for different  $b/a$ ratios. Two Gaussian processes, $X_{1}$ and $X_{2}$, with  time-delayed, Ornstein-Uhlenbeck  model---defined in~\eqref{eq:cross-o-u-delayed}, are simulated. It is assumed that $\sigma^{2}_{0}=1$, $\sigma^{2}_{1}=1.1$, $a=\sigma^{2}_{0}/\sigma^{2}_{1}$  and $\sigma_{2}^{2}= a^{2}\,\sigma_{1}^{2} (1 +b^{2}/a^{2})$.  Lower $b/a$ values imply higher cross correlation between $X_{1}$ and $X_{2}$. The time constants are $T=100$, $\tau_{1}=\tau_{2}=\tau_{0}=1$, while $\tau_{\ast}$ is determined from the ratio $\tau_{\ast}/\tau_{0}$. The continuous curves are based on the theoretical equilibrium estimates, i.e.,~\eqref{eq:tij-time-delay-expon} for $\delt < \tau_{\ast}$ (top and bottom panels), and~\eqref{eq:ifr-expon-small-delay} for $T_{1\to2}$ if $\delt \ge \tau_{\ast}$ (top). For every realization (pair  of length $N=1000$ time series) from an ensemble of $N_{\mathrm{sim}}=100$ simulations, the IFR is obtained using the data-driven estimator~\eqref{eq:tij}. Ensemble-based averages are  marked by stars, while the associated error bars  have a width of two standard deviations  (based on the ensemble estimates). }%
\label{fig:exponential-simulated}%
\end{figure}

Next, we use the  study design of Section~\ref{ssec:simul-gauss-time-delays}  to calculate the IFR for different combinations of  $\tau_{\ast}/\tau_{0}$ and  $b/a$ in the context of the linear regression model. The only difference is that herein the exponential model is used instead of the square exponential covariance in Section~\ref{ssec:simul-gauss-time-delays}.  The results are shown in  Fig.~\ref{fig:exponential-simulated}. The top panel shows $T_{1\to 2}$ and the bottom panel shows $T_{2\to 1}$.
The $T_{1\to 2}$  curves are calculated from~\eqref{eq:tij-time-delay-expon} for $\delt < \tau_{\ast}$ (i.e., for $\tau_{\ast}/\tau_{0}>0.1$) and from~\eqref{eq:ifr-expon-small-delay}  for $\delt \ge \tau_{\ast}$ ($\tau_{\ast}/\tau_{0} \le 0.1$). The $T_{2\to 1}$  curves are calculated from~\eqref{eq:tij-time-delay-expon} for all $\tau_{\ast}$. The $T_{1\to 2}$ curves show the same tendencies, albeit different shapes, with respect to $\tau_{\ast}/\tau_0$ and $b/a$ as the respective curves in Fig.~\ref{fig:gaussian-regression-simulated}.  The ensemble-based estimates are marked by stars with $\pm\,2\sigma_{\textrm{IFR}}$ error bars.  A small, yet systematic difference is observed between the theoretical and ensemble-based IFR values in the  $\delt \ge \tau_{\ast}$ (slow sampling) regime. This  is caused by  $\Oo\left((\delt - 2\tau_{\ast})^{2}\right)$ terms in the expansion~\eqref{eq:rhoij-taylor} for $\rho_{1,2}(\tau)$, which are not included in the leading-order theoretical estimate. The difference is more pronounced for smaller $b/a$ due to the amplification caused by the $(1-r^2)^{-1}$ factor for $r \approx \pm 1$.
The $T_{2\to1}$ are consistently close to the theoretical estimate (i.e., zero), while the dispersion is reduced for higher $\tau_{\ast}/\tau_{0}$ and $b/a$.

\section{Discussion}
\label{sec:discussion}

\subsection{IFR Interpretation}
In real-world problems, the absolute magnitude of the data-driven IFR (or its normalized counterpart~\cite{Liang15}) are used to determine causal relations, e.g.~\cite{Liang14,Stips16,dth19,Cong23}.  Values of $\lvert \hat{T}_{i\to j} \rvert \neq 0$  indicate  information flow in the direction $X_i \rightarrow X_j$.  The statistical significance of  non-zero $\hat{T}_{i\to j}$ estimates should be tested.  A parametric test based on  confidence intervals~\cite{Liang14}  is derived from estimates of the coefficients of  system~\eqref{eq:sode-motion-linear}.  A non-parametric test of $\hat{T}_{i\to j}$ uses random permutations ${X}^{(p)}_{i}$ of the series $X_{i}$ (i.e., the signal being investigated as potential driver) to obtain a baseline for non-significant deviations of $\hat{T}_{i\to j}$ from zero~\cite{dth19}.  The sign of $T_{i \to j}$  however, does not have a clear meaning. Our analysis provides a possible physical interpretation of the IFR sign.
 
More generally, we have shown that  the interpretation of $\hat{T}_{i\to j}$ for coupled stochastic processes depends on the smoothness (regularity) of the processes. For mean-square-differentiable processes with lead-lag correlations, the IFR is positive for  leading\textrightarrow lagging information flow and negative in the opposite direction.  If the processes are mean-square continuous but non-differentiable, the information flow does not share the same magnitude in both directions: the leading\textrightarrow lagging IFR is positive,  while the lagging\textrightarrow leading IFR can take zero or negative values.  For linear dynamical systems that satisfy~\eqref{eq:sode-motion-linear} with  diagonal diffusion matrix $\mathbf{B}$, the $\tij$ is exactly zero for any pair of indices $(i,j)$ such that $A_{j,i}=0$~\cite{Liang14}.  Inspection of~\eqref{eq:sode-motion-linear} shows that the differential of the stochastic Wiener process   forces discontinuous first-order derivatives for $\bfX(t)$.
Such  systems should be modeled with mean-square continuous (non-differentiable) stochastic processes  to avoid erroneous information flow estimates. For mean-square differentiable processes (such as the displacement from equilibrium  of coupled, linear, stochastic harmonic oscillators), the information flow  is non-zero in both directions due to the odd symmetry of the IFR (see Theorem~\ref{theorem:Tij-cs-msd}). In these cases,  analysis of the derivatives (i.e., displacement rates), which represent  mean-square continuous but non-differentiable processes, may be more revealing regarding the direction of information flow. 

To apply the results of our analysis to real-world  data, one needs information about the process regularity. According to Lemma~\ref{lemma:msd},  mean-square continuity and differentiability of  a stochastic process is determined by the regularity of its ACF at the origin. An optimal correlation model can be selected from the data using   likelihood-based~\cite{Brockwell09} or  Bayesian model selection, or the cross validation approach~\cite{Rasmussen06}.

Our analysis also suggests that in order to accurately identify the direction of information flow,  the sampling step should be smaller than the temporal delay and the latter be a fraction (optimally $\approx0.1$) of the correlation time. In addition, if the time delay $\tau_\ast$ of  interaction between processes is much smaller or much larger than the correlation time $\tau_0$,  the observed IFR tends to zero.  Nonetheless,  the IFR is able to detect information flow over a wide range of $\tau_{\ast}/\tau_{0}$ (cf. Figs.~\ref{fig:gaussian-regression-simulated} and~\ref{fig:exponential-simulated}).  In contrast,  WGC analysis requires specifying  an autoregressive order which should be equal or higher than the interaction time lag between the processes in order to  capture causal dependence~\cite{Bressler11,Stokes17}.  Estimates of $\tau_0$ and $\tau_\ast$ can be obtained by means of  model fitting and model selection approaches mentioned in the preceding  paragraph.

\subsection{Limitations of the analysis}
\label{ssec:limit}

Herein we focus on second-order ergodic (therefore, also stationary)  processes.  In practice, conditions of ergodicity require that time series be sufficiently long compared to the  correlation times of the processes involved. The regularity conditions (Lemma~\ref{lemma:msd}) do not require distributional assumptions for the stochastic processes.  However, for the numerical simulations we used Gaussian  processes because  (1)  ergodicity conditions can be established by means of Slutsky's theorem based exclusively on second-order moments and (2) the simulation of Gaussian time series is straightforward.

In many problems of interest (e.g., EEG recordings of brain activity at rest or during task execution),  the observed time series are non-stationary. Since the calculation of data-driven IFR estimates is straightforward, in non-stationary systems one can estimate  time-dependent IFR measures within contiguous time windows that are  locally stationary. In such cases, one can derive the information flow using  non-overlapping or overlapping time windows, e.g., as  in bootstrap subsampling~\cite{Politis94} and   windowed Fourier transforms~\cite{Press07}.
One of the goals of neuroscience is to identify  effective brain connectivity within and across different spectral activity bands~\cite{Cohen14}.  IFR has so far been used to investigate  brain connectivity  without spectral segmentation~\cite{dth19,Cong23}.  This question could be pursued by extending the results of Theorem~\ref{theor:IFR-ergodic-spectral}, using suitably modified IFR spectral expressions that involve band-limited spectral moments over different frequency bands of interest.

Liang has recently generalized the expression for the bivariate, data-driven IFR to multivariate systems containing $N \ge 2$ units (time series)~\cite{Liang21}. For $N=2$, this formulation  yields the same IFR between a pair of time series $x_{i}(t)$ and $x_{j}(t)$ with $i \neq j$ as the bivariate expression~\eqref{eq:tij}.  The multivariate expression also quantifies the self-interaction of each time series which is missing in~\eqref{eq:tij}.  Extension of the present analysis to the multivariate formulation is not straightforward, because the latter involves the cofactors of $N\times N$ covariance matrices.  In addition, verifying the admissibility of an $N$-variate cross-covariance model (where $N \gg 2$) requires controlling a large set of parameters.

Based on~\eqref{eq:tij}, IFR vanishes if $\hr_{i,j}=0$, i.e., if there are no cross correlations. This makes sense, because in the linear limit causation implies correlation (but not vice versa); hence, if two processes are causally linked,  their cross-correlation is expected to be non-zero.   A pathological case involves two deterministic cosine processes, one of which leads the other by $\pi/2$. Note that causality in this  system  is not detectable by the WGC method which is applicable to stochastic systems.  In spite of the  clear causal link, $\hr_{i,j}$ vanishes if it is estimated by  integrating over a multiple of the signals' period; therefore, IFR also vanishes.  However, it was recently shown that if  the cosine functions are  modeled as a stochastic bi-harmonic system with additive noise, the normalized IFR tends to unity as the noise variance tends to zero, thus recovering the causal relation~\cite{Liang21b}.  

One may ask if  IFR is related to other data-driven methods of causal inference such as WGC and transfer entropy~\cite{Schreiber00}.  The latter two  were shown to be entirely equivalent  under the Gaussian assumption~\cite{Barnett09}. To our knowledge, there are no systematic comparisons of IFR and WGC in the literature; a recent study focuses on differences in brain connectivity derived from the analysis of fMRI data~\cite{Cong23}.   The two methods have quite different origins.  WGC is based on a statistical test of the null hypothesis that the ``driver'' does not  impact the ``response'' in a statistically significant manner. WGC assumes that the system under study is described by means of a  linear, stationary, vector autoregressive model (although there are extensions that relax these assumptions~\cite{Bressler11,Marinazzo11}).

Liang’s IFR is obtained by calculating the flow of information (expressed in terms of entropy) between subspaces of a stochastic, nonlinear dynamical system. This leads to IFR equations  that involve multidimensional integrals of joint probability density functions which are not in general amenable to explicit solution.  The data-driven IFR~\eqref{eq:tij}, studied herein, was derived by applying maximum likelihood estimation to time series data, assuming a linear stochastic system that satisfies~\eqref{eq:sode-motion-linear}~\cite{Liang14}. Nonetheless, the data-driven IFR has  been successfully applied to  benchmark dynamical systems such as 
the Baker,  H\'{e}non and Kaplan–Yorke maps, and  R\"{o}ssler oscillators~\cite{Liang16,Liang21,Liang18}.    A deeper physical understanding of IFR's magnitude and sign is needed for nonlinear systems. In the linear case, it remains to be investigated if there is a deeper connection between IFR and WGC.  IFR has a practical advantage over WGC, because IFR's computational complexity is  controlled by the calculation of correlations for two time lags (zero and $\delt$)---therefore it scales linearly with the size of the time series---in contrast with the cubic scaling of WGC (cf. Section~\ref{sec:intro}).

\section{Conclusion}
\label{sec:conclusions}
We investigated the data-driven Liang information flow rate between coupled, second-order ergodic stochastic processes. We defined the equilibrium IFR  (Theorem~\ref{theor:IFR-ergodic}), and  we developed a spectral formulation for the equilibrium IFR in terms of spectral moments of the coupled processes  (Theorem~\ref{theor:IFR-ergodic-spectral}).  We showed that the continuous sampling limit ($\delt \to 0$)  of the equilibrium IFR can be defined for both mean-square differentiable (Theorem~\ref{theorem:Tij-cs-msd}) and mean-square continuous (Theorem~\ref{theorem:Tij-cs-msc-delay}) processes. We also derived  leading-order finite-step corrections in the case of mean-square differentiable processes (Corollary~\ref{corol:finite-step-msd}).
Furthermore, we established that the equilibrium IFR vanishes for separable cross-correlation models $\bfC(\tau)= \mathbf{c} \rho(\tau)$, where $\mathbf{c}$ is a $D\times D$ covariance matrix and $\rho(\tau)$ an admissible scalar correlation function. This result holds for second-order  ergodic processes independently of  regularity properties (Proposition~\ref{propo:separ-zero-ifr}).

For  mean-square differentiable processes with non-separable covariance kernels, we found that  the IFR  has the same magnitude in both directions but  is positive in one direction and negative  in the other   (Theorem~\ref{theorem:Tij-cs-msd}).  We also investigated the equilibrium IFR  for cross-correlation models featuring time delays. In the case of differentiable covariance kernels, the IFR exhibits an odd symmetry: it is positive in the direction from the leading to the lagging series and negative in the opposite direction  (Theorem~\ref{theorem:Tij-cs-msd-delay}). These general results were  explored   by means of a simple regression model (Section~\ref{ssec:msd-regression-delayed}). The IFR antisymmetry  is broken for mean-square continuous (but non-differentiable) processes (Theorem~\ref{theorem:Tij-cs-msc-delay}). For the Ornstein-Uhlenbeck correlation model, we have shown that the IFR from the leading to the lagging series is positive, while the IFR from lagging to the leading series vanishes provided that the characteristic time constants of the ACFs and CCFs match (Remark~\ref{rem:asymmetric-tij-msc}).  Finally, we showed that for mean-square differentiable processes, as well as mean-square continuous processes with sampling step smaller than the time delay ($\delt < \tau_{\ast}$), the leading-order  equilibrium IFR is  independent of $\delt$. In the case of mean-square continuous processes with $\tau_{\ast}\le \delt$ (slow sampling regime), the leading-order term  is $\Oo(1/\delt)$.

\appendices
\section{Proof of Theorem~\ref{theorem:Tij-cs-msd}: IFR continuous sampling limit}
\label{app:Tij-differ}

\begin{IEEEproof}
(1) According to~\eqref{eq:E-Tij}, $\tij(\delt)$ involves the product of  $\rho_{i,j}(0)/[1-\rho^{2}_{i,j}(0)]$, which is independent of $\delt$, with $r_{i,dj}(\delt) - \rho_{i,j}(0)\, r_{j,dj}(\delt)$. Hence, to determine  $\lim_{\delt \to 0}\tij(\delt)$,  the limit $r_{i,dj}(\delt)$ as $\delt \to 0$ needs to be evaluated, where $r_{i,dj}(\delt)$ is defined in~\eqref{eq:ridj-ergodic}.  Assuming mean-square differentiability, the limit $\delt \to 0$ becomes
\beq
\label{eq:limit-ridj}
\lim_{\delt \to 0} r_{i,dj}(\delt) = \lim_{\delt \to 0}\frac{1}{\delt}\left[ \rho_{i,j}(\delt) - \rho_{i,j}(0)\right].
\eeq

\noindent We consider the terms $r_{i,dj} \, (i \neq j)$ and $r_{j,dj}$  separately.

{Term 1} ($r_{i,dj}$ for $i \neq j$):
The  CCF  $\rho_{i,j}(\tau)$
does not necessarily peak at $\tau=0$; in fact, the peak appears at $\tau \neq 0$ if the  influence of $X_{i}(t)$ on $X_{j}(t)$ occurs after a finite time delay. Then, the first-order derivative of $\rho_{i,j}(\tau)$ at zero does not vanish. The Taylor expansion of $\rho_{i,j}(\delt)$ around $\delt =0$ leads to
\beq
\label{eq:rho-taylor}
\rho_{i,j}(\delt) = \rho_{i,j}(0) + \rho_{i,j}^{(1)}(0) \, \delt + \frac{\rho_{i,j}^{(2)}(0)}{2} \, \delt^2 + \Oo (\delt^3).
\eeq
It follows from~\eqref{eq:limit-ridj} and~\eqref{eq:rho-taylor} that  $\lim_{\delt \to 0} r_{i,dj}(\delt)=  \rho_{i,j}^{(1)}(0)$ for $i \neq j$.

{Term 2} ($r_{i,dj}$ for $i=j$): According to Lemma~\ref{lemma:msd} it holds that $\rho_{j,j}^{(1)}(0)=0$. Since $\rho_{j,j}(0)=1$,  $\rho_{j,j}(\delt)$ admits the following Taylor expansion around $\delt=0$:
\beq
\label{eq:rho-taylor-jj}
\rho_{j,j}(\delt) = 1 + \frac{1}{2} \rho_{j,j}^{(2)}(0) \, \delt^2 + \Oo (\delt^4).
\eeq
Based on the above expansion, the limit~\eqref{eq:limit-ridj}   is given by
\[
\lim_{\delt \to 0} r_{j,dj}(\delt) = \lim_{\delt \to 0} \left[ \frac{\delt}{2}\, \rho_{j,j}^{(2)}(0) + \Oo (\delt^3) \right] = 0.
\]
Therefore, the term $\propto r_{j,dj}(\delt)$ in~\eqref{eq:E-Tij} vanishes at the limit $\delt \to 0$.  This result signifies that differentiable processes $X(t)$ are uncorrelated  with $\dot{X}(t)$.
Hence, only   $\lim_{\delt \to 0}r_{i,dj}(\delt)=\rho^{(1)}_{i,j}(0)$ for $i \neq j$ enters in $\tijz$ leading to~\eqref{eq:tij-differentiable}.  

According to~\eqref{eq:tij-differentiable}, the sign of IFR is determined by the sign of the  product $\rho_{i,j}(0) \rho_{i,j}^{(1)}(0)$.  The latter is  the first-order derivative of $\rho_{i,j}^{2}(\tau)$ with respect to $\tau$ evaluated at $\tau=0$.

\smallskip

(2) The antisymmetry of $\tijz$ follows from~\eqref{eq:tij-differentiable}.  Since $\rho_{i,j}(0)=\rho_{j,i}(0)$, the sign of $\tjiz$ is determined from $\rho^{(1)}_{j,i}(0)$.
CCFs respect the \emph{reflection symmetry}  $\rho_{i,j}(\delt)=\rho_{j,i}(-\delt)$~\cite{Papoulis02}. Using the Taylor series expansion of both terms around $\delt=0$  and equating terms of the same order in $\delt$, it follows that the $n$-order derivative satisfies $\rho_{j,i}^{(n)}(0) = (-1)^{n} \,\rho_{i,j}^{(n)}(0)$ and thus  $\rho_{j,i}^{(1)}(0) = -\,\rho_{i,j}^{(1)}(0)$. Hence, $\tjiz= - \tijz$. 
\end{IEEEproof}

\section{Proof of Corollary~\ref{corol:finite-step-msd}: Finite-time-step corrections}
\label{app:finite-step-msd}

\begin{IEEEproof}
We use the equilibrium IFR~\eqref{eq:equilibrium-ifr} and define $g_{i,j}(\delt) \triangleq \frac{1}{\delt}\left[\rho_{i,j}(\delt) - \rho_{i,j}(0)\rho_{j,j}(\delt)\right]$.  Then~\eqref{eq:equilibrium-ifr} is expressed as
\beq
\label{eq:tij-gij}
\tij(\delt) = \frac{\rho_{i,j}(0)}{1 - \rho_{i,j}^{2}(0)}\,g_{i,j}(\delt)\,.
\eeq
Using~\eqref{eq:rho-taylor}-\eqref{eq:rho-taylor-jj} and the above  definition of $g_{i,j}(\delt)$, the following Taylor series expansion is obtained for $g_{i,j}(\delt)$
\beq
\label{eq:gij-dt}
g_{i,j}(\delt) = \rho_{i,j}^{(1)}(0)+ \frac{\delt}{2}\left[\rho_{i,j}^{(2)}(0) - \rho_{i,j}(0)\rho_{j,j}^{(2)}(0) \right] + \Oo(\delt^2)\,.
\eeq

Case 1: $\rho_{i,j}^{(1)}(0) = 0$.  If this condition holds, according to~\eqref{eq:tij-differentiable} the continuous sampling IFR vanishes, i.e. $\tijz=0$. By setting $\rho_{i,j}^{(1)}(0) = 0$ in ~\eqref{eq:gij-dt} we obtain from~\eqref{eq:tij-gij} the second branch of~\eqref{eq:dtij-msd}.

Case 2: $\rho_{i,j}^{(1)}(0) \neq 0$.  In this case~\eqref{eq:gij-dt} leads to
$g_{i,j}(\delt) = \rho_{i,j}^{(1)}(0)+ \Oo(\delt)/2$. In addition, $\tijz \neq 0$ according to~\eqref{eq:tij-differentiable}.  The leading correction is thus given by
$\delta \tijz = \tij(\delt) -  \tijz - \Oo(\delt^2)$.
Based on~\eqref{eq:tij-gij} and~\eqref{eq:gij-dt} we obtain
\[
\delta \tijz  = \frac{\rho_{i,j}(0)}{1 - \rho_{i,j}^{2}(0)}\,
\frac{\delt\left[\rho_{i,j}^{(2)}(0) - \rho_{i,j}(0)\rho_{j,j}^{(2)}(0) \right]}{2}\,.
\]
Finally, if we multiply and divide the right-hand side of the above with $\rho_{i,j}^{(1)}(0) \neq 0$ and recall~\eqref{eq:tij-differentiable} for $\tijz$, the first branch of~\eqref{eq:dtij-msd} is recovered.

\end{IEEEproof}

\section{Proof of Theorem~\ref{theorem:Tij-cs-msc-delay}: Mean-square  continuous processes with time-delayed correlations}
\label{app:ms-continuous-delay}

\begin{IEEEproof}
For $i \neq j$, $\rho_{i,j}(0)=C_{0}(-\epsilon_{i,j}\tau_{\ast})/\sigma_{i}\sigma_{j}$; given the symmetry of  $C_{0}(\cdot)$ it holds that $C_{0}(-\epsilon_{i,j}\tau_{\ast})=C_{0}(\tau_{\ast})$ and thus  $\rho_{i,j}(0) =r$.
We use a unified notation for  $i=j$ and $i \neq j$, by introducing  functions $\phi_{i,j}(u): \R_{\ge 0} \to \R$ such that $\rho_{i,j}(\tau)=\phi_{i,j}( \lvert \tau-\tilde{\tau} \rvert)$; for $i=j$, $\phi_{i,i}(u)=\rho_{i,i}(\tau)$, where $u=\lvert \tau-\tilde{\tau} \rvert$, while $\phi_{i,j}(u)=C_{0}(\tau)/\sigma_{i}\sigma_{j}$ for $i \neq j$.  The temporal offset  $\tilde{\tau} \in \R$ is given by  $\tilde{\tau}=\epsilon_{i,j} \tau_{\ast}$ so that it produces the correct sign for the leading/lagging series and  vanishes for $i=j$. 

Based on the conditions specified above for $C_0(\cdot)$, the functions  $\phi_{i,j}(u)$
are differentiable for all $u \in \R$ and have a global maximum at $u=0$.  For  the Ornstein-Uhlenbeck model all of these functions are of the form $\varphi(u)=\exp(-u/\tau_0)$; a Taylor expansion of $\varphi(\cdot)$  around
$\tau=0_{+}$ (thus $u=\lvert \tilde{\tau} \rvert$) yields
\begin{align}
\label{eq:taylor-phi}
 \varphi( \lvert \tau-\tilde{\tau} \rvert ) = &  \varphi(\lvert \tilde{\tau} \rvert) +  \varphi^{(1)}(\lvert \tilde{\tau} \rvert)  \left( \lvert \,\tau-\tilde{\tau} \rvert - \lvert \tilde{\tau} \rvert \,\right)
 \nonumber \\
 & + \Oo\left( \, \lvert \,\tau-\tilde{\tau} \rvert - \lvert \tilde{\tau} \rvert\,\right)^2 \,,
\end{align}
where $\varphi^{(1)}(\lvert \tilde{\tau} \rvert) \triangleq
\left. \varphi^{(1)}(u) \right|_{u=\lvert \tilde{\tau} \rvert}$; thus, if $\tilde{\tau}\to 0$  the derivative $\varphi^{(1)}(0_{+})$ is evaluated.
For example, the following first-order Taylor approximation is easily confirmed for $\lvert \tau-\tilde{\tau} \rvert \ll \tau_{0}$: 
\[
\exo^{ -\lvert \tau-\tilde{\tau} \rvert/\tau_{0}} \approx \exo^{ -\tilde{\tau}/\tau_0} \left( 1 - \frac{\lvert \tau-\tilde{\tau} \rvert}{\tau_{0}} - \frac{\lvert \tilde{\tau} \rvert}{\tau_{0}}\right)\,.
\]

In light of~\eqref{eq:taylor-phi} and recalling that $\tilde{\tau}= \epsilon_{i,j}\tau_{\ast}$, the Taylor expansion for  the \emph{forward finite difference}  $\rho_{i,j}(\delt)$  around $\delt=0$ is as follows:
\begin{align}
\label{eq:rhoij-taylor}
\rho_{i,j}(\delt) = & \rho_{i,j}(0) + \rho_{i,j}^{(1)}(0_{+}) \, \left( \vert \delt -\epsilon_{i,j}\tau_{\ast}\rvert - \lvert \epsilon_{i,j}\tau_{\ast}\rvert \right)
\nonumber \\
+ &  \frac{1}{2} \rho_{i,j}^{(2)}(0_{+}) \, \left( \vert \delt -\epsilon_{i,j}\tau_{\ast}\rvert - \lvert \epsilon_{i,j}\tau_{\ast}\rvert \right)^2 + \Oo (\delt^3),
\end{align}
\noi where $\rho_{i,j}^{(n)}(0_{+})=\phi^{(n)}_{i,j}(\lvert \tilde{\tau}\rvert)$ is the limit of the $n$th-order derivative of  $\phi_{i,j}(u)$ with respect to $u$ as $\delt \to 0_{+}$.

Based on the equilibrium IFR expression~\eqref{eq:equilibrium-ifr} and the Taylor expansion~\eqref{eq:rhoij-taylor}, and taking into account that $\lvert \epsilon_{i,j}\tau_{\ast}\rvert = \tau_{\ast}$, the leading-order in $\delt$ IFR approximation for $i \neq j$ is given by
\begin{align}
\label{eq:equil-ifr-msc}
\tij(\delt) = & \frac{r \, }{1 - r^{2}}
\, \frac{1}{\delt}
\left[ \,\rho_{i,j}^{(1)}(0_{+}) \left( \,\vert \delt -\epsilon_{i,j}\tau_{\ast}\rvert -\tau_{\ast} \,\right)  \right.
\nonumber \\
\quad \quad \quad \quad & - \left.  \rho_{i,j}(0)\, \rho_{j,j}^{(1)}(0_{+}) \delt\,\right]\,.
\end{align}

\noindent Based on~\eqref{eq:equil-ifr-msc} and $\epsilon_{2,1}=-1$,  the leading-order $T_{2 \to 1}(\delt)$ approximation for information flow in the receiver$\rightarrow$driver ($2\to 1$) direction becomes
\[
\label{eq:t21-delay-msc}
T_{2\to 1}(\delt)= \frac{r}{1-r^{2}}\left[ \,\rho_{2,1}^{(1)}(0_{+})  - r\, \rho_{1,1}^{(1)}(0_{+}) \,\right]\,.
\]
Hence, $T_{2 \to 1}(\delt)$ is independent of $\delt$ and the continuous-sampling limit $\mathcal{T}_{2\to1}$ is well defined. This proves the expression~\eqref{eq:tij-cs-lim-continuous-delay} for receiver$\rightarrow$driver IFR as well as statement (3).

The expression~\eqref{eq:equil-ifr-msc} for $T_{1 \to 2}(\delt)$ involves  $\epsilon_{1,2}=1$. For information flow in the driver$\rightarrow$receiver ($1\to 2$)  direction, we need to consider two cases: (1) for $\delt > \tau_{\ast}$ we obtain
\begin{subequations}
\label{eq:t12-delay-msc}
\begin{align}
 T_{1\to 2}(\delt)=
  \frac{r}{1-r^{2}}\left[ \rho_{1,2}^{(1)}(0_{+}) \left(1 - \frac{2\tau_\ast}{\delt} \right) - r\, \rho_{2,2}^{(1)}(0_{+}) \right],
\end{align}
and (2) for $\delt \le \tau_{\ast}$
\begin{align}
 T_{1\to 2}(\delt)=    \frac{r}{1-r^{2}}\left[ -\rho_{1,2}^{(1)}(0_{+})  - r\, \rho_{2,2}^{(1)}(0_{+}) \right]  \,  \,.
\end{align}
\end{subequations}

\medskip
The second branch of~\eqref{eq:t12-delay-msc} proves the continuous sampling limit $\mathcal{T}_{1\to 2}$~\eqref{eq:tij-cs-lim-continuous-delay}  for IFR in the driver$\rightarrow$receiver direction (statement 1). It also proves statement (4) in Theorem~\ref{theorem:Tij-cs-msc-delay} which applies if the sampling step exceeds the delay.


To prove~\eqref{eq:antisymmetry}, i.e., statement (2) in Theorem~\ref{theorem:Tij-cs-msc-delay}, we evaluate the sum $\tijz+ \tjiz$ using~\eqref{eq:equil-ifr-msc}.  We employ the symmetry  $C_{0}(\tau_\ast)=C_{0}(-\tau_\ast)$, and the fact that  $C^{(1)}(u)$ is an odd function of $u$, i.e., $C^{(1)}(u)=-C^{(1)}(-u)$.    The latter implies that 
$\rho^{(1)}_{i,j}(0_{+})+\rho^{(1)}_{j,i}(0_{+})$ vanishes in  $\tijz + \mathcal{T}_{j \to i}$ 
(due to the odd symmetry of the first derivative of the CCF, cf. Appendix~\ref{app:Tij-differ}).  Thus, the expression~\eqref{eq:antisymmetry} follows by adding the second terms on the right-hand side of~\eqref{eq:equil-ifr-msc}.
Finally, the non-negativity of   $\tijz + \mathcal{T}_{j \to i}$ is due to the fact that $\rho_{j,j}^{(1)}(0_{+})<0$ and $0 \le r^{2} \le 1$.
\end{IEEEproof}

\section*{Acknowledgment}
We acknowledge constructive email communications with X. San Liang (Fudan University and Southern Marine Science and Engineering Guangdong Laboratory, China) and Arif Babul (University of Victoria, Canada).

\ifCLASSOPTIONcaptionsoff
  \newpage
\fi



\end{document}